\documentclass[useAMS]{mn2e}
\usepackage{graphicx,amssymb,amsmath}
\usepackage{hyperref,xcolor}
\usepackage{subfigure}
\usepackage{pdflscape}
\usepackage{mathrsfs}

\newcommand{\scs}{\scriptsize}

\title[The evolution of the Milky Way from open clusters] {The evolution of the Milky Way: New insights from open clusters }
\author[A. B. S. Reddy, D. L. Lambert and S. Giridhar]
  { Arumalla B. S. Reddy$^{1}$\thanks{E-mail: bala@astro.as.utexas.edu},
   David L. Lambert$^1$ and Sunetra Giridhar$^2$  \\
   $^1$W.J. McDonald Observatory and Department of Astronomy, The University of Texas at Austin, Austin, TX 78712, USA \\
   $^2$Indian Institute of Astrophysics, Bangalore 560034, India }

\begin{document}

\date{Accepted 2016 September 07. Received 2016 September 07; in original form 2016 May 12}

\pagerange{\pageref{firstpage}--\pageref{lastpage}} \pubyear{2016}

\maketitle

\label{firstpage}

\begin{abstract}

We have collected high-dispersion echelle spectra of red giant members in the twelve open clusters (OCs) and derived stellar parameters and chemical abundances for 26 species by either line equivalent widths or synthetic spectrum analyses. 
We confirm the lack of an age$-$metallicity relation for OCs but argue that such a lack of trend for OCs arise from the limited coverage in metallicity compared to that of field stars which span a wide range in metallicity and age. We confirm that the radial metallicity gradient of OCs is steeper (flatter) for R$_{\rm gc}<\,$12 kpc ($>$12 kpc). We demonstrate that the sample of clusters constituting a steep radial metallicity gradient of slope $-$0.052$\pm$0.011 dex kpc$^{-1}$ at R$_{\rm gc}<$ 12 kpc are younger than 1.5 Gyr and located close to the Galactic midplane ($\lvert\,z\rvert<\,$0.5 kpc) with kinematics typical of the thin disc. Whereas the clusters describing a shallow slope of $-$0.015$\pm$0.007 dex kpc$^{-1}$ at R$_{\rm gc}>$ 12 kpc are relatively old, thick disc members with a striking spread in age and height above the midplane (0.5$\,<\lvert\,z\rvert<\,$2.5 kpc). Our investigation reveals that the OCs and field stars yield consistent radial metallicity gradients if the comparison is limited to samples drawn from the similar vertical heights. We argue via the computation of Galactic orbits that all the outer disc clusters were actually born inward of 12 kpc but the orbital eccentricity has taken them to present locations very far from their birthplaces.

\end{abstract}


\section{Introduction} 
The formation and evolution of galaxies is one of the major puzzles of astrophysics. Large-scale cosmological simulations utilizing the currently favoured Lambda Cold Dark Matter ($\Lambda$CDM) paradigm of cosmology predicts the large-scale distribution of galaxies in space as observed today, but we have not yet met with equal success on small scales due to incomplete knowledge of how the baryons are distributed (Vogelsberger et al. 2014). As these baryons mostly reside in stellar discs in galaxies, understanding the formation and evolution of galactic discs is the heart of the galaxy formation theory (Feltzing 2015 and references therein). 

In this regard, the Milky Way Galaxy provides an excellent testing ground for models of galaxy formation and evolution thanks to the ability to resolve individual stars of its stellar populations and analyse them in exquisite detail. As the disc formed dissipatively and evolved dynamically, much of the dynamical information is lost. However, the chemical content of the disc stars may have preserved the dissipative history, the key to unraveling the formation history of the Milky Way (De Silva et al. 2009; Freeman \& Bland-Hawthorn 2002). Since the surface density of gas and star formation rate (SFR) in the Galactic disc have varied, as seen in many galaxies, the measured chemical abundances are a function of position as well. Therefore, measurement of the chemical abundance distribution in the disc (i.e. the radial abundance gradient) and the gradient's temporal variation over the disc's lifetime present strong observational constraints on the theoretical models of Galactic chemical evolution (GCE; Magrini et al. 2009; Minchev et al. 2013; Kubryk et al. 2015).

Abundance gradients in the Milky Way can be measured using a wide variety of tracers including the young population of H\,{\sc ii} regions, OB stars, and Cepheid variables, and planetary nebulae of different ages, cool unevolved stars and red giants in the field (see for example, Magrini et al. 2010; Cheng et al. 2012; Hayden et al. 2014) and in open clusters (OCs; Friel et al. 2002, 2010; Magrini et al. 2010; Pancino et al. 2010; Yong et al. 2012;  Frinchaboy et al. 2013). The cooler main sequence stars or red giants which are members of OCs provide not only abundance estimates for elements sampling all the major processes of stellar nucleosynthesis -- but a collection of stars whose age, distance, kinematics and metallicity can be measured with greater certainty than for field stars. Moreover, it is possible from an OC's space motion and a model of the Galactic gravitational potential to study the dynamics and estimate a cluster's  birthplace (Wu et al. 2009; Vande Putte et al. 2010).

In recent years, modern high-resolution spectrographs and large reflectors have provided high-quality spectra of stars in OCs for secure measures of abundances, and extended abundance estimates to distant OCs in the direction of Galactic anti-centre (Carraro et al. 2007; Yong et al. 2005, 2012; Sestito et al. 2006, 2008). However, not all the investigators have arrived at similar measures of the radial metallicity gradient which may be partly related to the different methods adopted (see for example, Magrini et al. 2009; Heiter et al. 2014). Our current understanding suggests a gradient of about $-$0.06 dex kpc$^{-1}$ (Friel et al. 2002; Pancino et al. 2010) to $-$0.20 dex kpc$^{-1}$ (Frinchaboy et al. 2013) in the radial range 5 to 10 kpc with a nearly flat trend ($-$0.02 dex kpc$^{-1}$) beyond 13 kpc (Carraro et al. 2007; Sestito et al. 2008; Pancino et al. 2010; Yong et al. 2012; Frinchaboy et al. 2013; Cantat-Gaudin et al. 2016). 

It is especially puzzling that the flattening of the metallicity distribution of OCs at large radii is not shown by the Galactic field stars (Cheng et al. 2012; Hayden et al. 2014) including Cepheids (Luck \& Lambert 2011; Genovali et al. 2014) where field stars suggest a constant steep decline of metallicity out to R$_{\rm gc}$ of 18 kpc. Another fascinating result, in common for both the field stars and OCs at large radii, is the evidence for enhanced [$\alpha$/Fe] ratios in the outer Galactic disc (Yong et al. 2005; Bensby et al. 2011; Luck \& Lambert 2011). The fact that the $\alpha$-enriched clusters older than 4$-$5 Gyr, reminiscent of rapid star formation history, have populated the Galactic disc beyond 12 kpc, where the star formation is slow due to low surface density of gas, is a puzzle in conflict with the inside-out formation scenarios of the Milky Way (Bovy et al. 2012; Ro\v{s}kar et al. 2013). A significant flattening of the metallicity gradient observed in the outskirts of other spiral galaxies (Scarano \& L\'{e}pine 2013) calls for a thorough understanding of many processes regulating the evolution of Milky Way.

This paper extends our homogeneous abundance analysis of OCs to a new sample of OCs to aid in exploring further the evolution of the Galactic disc.
This is our fourth paper reporting a comprehensive abundance measurements for red giants in twelve OCs in the radial 7.7 to 9.7 kpc with ages 71 to 730 Myr lacking detailed information on their chemical composition. The clusters selected for abundance analysis included: NGC 1647, NGC 1664, NGC 2099, NGC 2281, NGC 2287, NGC 2345, NGC 2437, NGC 2548, NGC 2632, NGC 6633, NGC 6940, and NGC 7209. All these clusters save NGC 2099, NGC 2287, NGC 2632 and NGC 6633 are analysed for the first time. The chemical content of NGC 2099 was previously measured for the elements Na, Al, $\alpha$-elements (Mg, Si, Ca, and Ti), iron-peak elements (Sc, V, Cr, Fe, Co, and Ni), and the $s$-process elements (Y, Ba, La, and Nd) by Pancino et al. (2010). The chemical composition of NGC 2632 was estimated previously for the elements Na, Al, Si, Ca, Ti, Cr, Fe and Ni by Pace, Pasquini \& Fran{\c c}ois (2008), and for 17 elements in the range from O to Ba by Yang, Chen \& Zhao (2015). The chemical abundances of none of the elements but for iron was explored previously for red giants in NGC 2287 and NGC 6633 by Santos et al. (2012) and Santos et al. (2009), respectively. For NGC 2099 and NGC 2632, we either provide a more extensive analysis of stars previously analysed or the first analysis of a member star while NGC 2287 and NGC 6633 are analysed comprehensively for the first time for many elements. 
 
We have presented in our previous papers (Reddy et al. 2012, 2013, 2015) abundance measurements for 16 OCs with ages from 130 Myr to 4.3 Gyr and R$_{\rm gc}$s from 8.3 and 11.3 kpc. We combine our sample of twenty-eight OCs (12 from this study and 16 from our previous papers) with a sample of 51 OCs drawn from the literature for which we have remeasured the chemical content using their published EWs, our models, linelists and reference solar abundances to establish a common abundance scale. Useful data on chemical composition of the literature sample is presented previously in Table 13 from Reddy et al. (2015).

This paper is organized as follows: In Section 2 we describe observations, data reduction and radial velocity measurements. Section 3 is devoted to the abundance analysis and Section 4 to discussing the age$-$abundance relations. We present in Section 5 the radial abundance distribution of OCs and field stars and discuss the results in comparison with theoretical chemodynamical models of GCE. In Section 6 we compute the birthplace of OCs with other relevant orbital parameters and focus on discussing the origin of older OCs in the outer Galactic disc. Finally, Section 7 provides the conclusions.

\begin{table*}
\centering
\caption{The journal of the observations for each of the cluster members analysed in this paper.} 
\label{log_observations}
\begin{tabular}{lccccccccclc}   \hline
\multicolumn{1}{l}{Cluster}& \multicolumn{1}{c}{Star} & \multicolumn{1}{c}{$\alpha(2000.0)$}& \multicolumn{1}{c}{$\delta(2000.0)$}&  
\multicolumn{1}{c}{V}& \multicolumn{1}{c}{B-V} & \multicolumn{1}{c}{V-K$_{\rm s}$} &  \multicolumn{1}{c}{J-K$_{\rm s}$} &
\multicolumn{1}{c}{$RV_{\rm helio}$} & \multicolumn{1}{l}{S/N at} & \multicolumn{1}{l}{Date of} & \multicolumn{1}{c}{Exp. time}  \\
\multicolumn{1}{c}{}& \multicolumn{1}{c}{}& \multicolumn{1}{c}{(hh mm ss)}& \multicolumn{1}{c}{($\degr$ $\arcmin$ $\arcsec$)}&
\multicolumn{1}{c}{(mag)}& \multicolumn{1}{c}{ } & \multicolumn{1}{c}{ }& \multicolumn{1}{c}{ } & 
\multicolumn{1}{c}{(km s$^{-1}$)} & \multicolumn{1}{l}{6000 \AA } & \multicolumn{1}{l}{observation} & \multicolumn{1}{c}{(sec)}  \\
\hline
 NGC 1647 &  2  & 04 46 05.10 & $+$18 48 02.67 & 07.47 &$+$1.50 &$+$3.59 &$+$0.84 &$-$07.8$\pm$0.2 & 300 & 25-12-2013 & 2$\times$1200 \\
          & 105 & 04 46 35.90 & $+$19 29 39.32 & 08.45 &$+$1.60 &$+$4.00 &$+$0.98 &$-$07.8$\pm$0.2 & 190 & 25-12-2013 & 2$\times$1800 \\
 NGC 1664 &  17 & 04 51 02.11 & $+$43 38 45.70 & 11.41 &$+$1.08 &$+$2.67 &$+$0.67 &$+$09.2$\pm$0.1 & 150 & 25-12-2013 & 3$\times$1800 \\
          &  75 & 04 51 19.72 & $+$43 42 16.54 & 11.06 &$+$0.95 &$+$2.44 &$+$0.67 &$+$06.4$\pm$0.3 & 100 & 24-12-2013 & 3$\times$1800 \\
 NGC 2099 &  34 & 05 52 15.10 & $+$32 31 40.92 & 10.90 &$+$1.14 &$+$2.81 &$+$0.69 &$+$10.3$\pm$0.3 & 120 & 21-11-2013 & 2$\times$1800 \\
          &  64 & 05 52 13.04 & $+$32 34 16.60 & 11.15 &$+$1.36 &$+$2.97 &$+$0.76 &$+$08.6$\pm$0.2 & 160 & 21-11-2013 & 2$\times$1800 \\
 NGC 2281 &  55 & 06 48 15.09 & $+$41 04 22.23 & 08.86 &$+$0.96 &$+$2.27 &$+$0.57 &$+$19.6$\pm$0.2 & 150 & 27-12-2013 & 1$\times$1200 \\
          &  63 & 06 48 21.72 & $+$41 18 08.36 & 07.24 &$+$1.37 &$+$3.06 &$+$0.83 &$+$19.3$\pm$0.2 & 280 & 27-12-2013 & 1$\times$1200 \\
 NGC 2287 &  75 & 06 45 43.01 & $-$20 51 09.59 & 07.43 &$+$1.28 &$+$2.82 &$+$0.78 &$+$24.0$\pm$0.1 & 400 & 20-11-2013 & 1$\times$1800 \\
          &  97$^{a}$ & 06 46 04.84 & $-$20 36 24.91 & 07.80 &$+$1.16 &$+$2.59 &$+$0.70 &$+$22.9$\pm$0.1 & 370 & 20-11-2013 & 1$\times$1800 \\
          & 107$^{a}$ & 06 46 33.28 & $-$20 48 42.63 & 07.79 &$+$1.15 &$+$2.59 &$+$0.66 &$+$26.3$\pm$0.1 & 370 & 20-11-2013 & 1$\times$1800  \\
 NGC 2345 &  34$^{a}$ & 07 08 21.84 & $-$13 10 23.25 & 09.94 &$+$1.50 &$+$4.21 &$+$0.97 &$+$63.6$\pm$0.4 & 120 & 25-12-2013 & 2$\times$1800 \\
          &  43 & 07 08 26.32 & $-$13 11 14.44 & 10.70 &$+$1.81 &$+$4.54 &$+$1.09 &$+$58.0$\pm$0.4 & 180 & 27-12-2013 & 2$\times$1800 \\
          &  60 & 07 08 30.37 & $-$13 13 52.54 & 10.48 &$+$1.82 &$+$4.51 &$+$1.11 &$+$57.4$\pm$0.3 & 160 & 25-12-2013 & 2$\times$1800 \\
 NGC 2437 &  29$^{a}$ & 07 41 37.61 & $-$14 43 12.98 & 10.86 &$+$1.19 &$+$2.92 &$+$0.76 &$+$52.7$\pm$0.2 & 190 & 27-12-2013 & 2$\times$1800 \\
          & 174$^{a}$ & 07 41 51.52 & $-$14 54 29.28 & 10.70 &$+$1.11 &$+$2.50 &$+$0.60 &$+$45.8$\pm$0.2 & 160 & 27-12-2013 & 2$\times$1800 \\
          & 242$^{a}$ & 07 41 19.43 & $-$14 48 47.45 & 10.19 &$+$1.12 &$+$2.74 &$+$0.69 &$+$57.2$\pm$0.2 & 160 & 25-12-2013 & 2$\times$1800 \\
 NGC 2548& 1218 & 08 13 35.43 & $-$05 53 02.18 & 09.64 &$+$0.94 &$+$2.18 &$+$0.52 &$+$08.7$\pm$0.2 & 110 & 24-12-2013 & 2$\times$1200 \\
         & 1296$^{a}$ & 08 13 44.83 & $-$05 48 00.89 & 09.27 &$+$0.77 &$+$2.11 &$+$0.58 &$+$13.7$\pm$0.2 & 180 & 25-12-2013 & 2$\times$1500 \\
         & 1560$^{a}$ & 08 14 17.03 & $-$05 54 00.62 & 08.19 &$+$1.10 &$+$2.77 &$+$0.73 &$-$00.5$\pm$0.2 & 200 & 25-12-2013 & 2$\times$1200 \\
         & 1628 & 08 14 28.12 & $-$05 42 16.14 & 09.47 &$+$1.02 &$+$2.39 &$+$0.62 &$+$08.0$\pm$0.2 & 240 & 17-11-2011 & 2$\times$1800 \\
 NGC 2632 & 212 & 08 39 50.71 & $+$19 32 26.93 & 06.60 &$+$0.96 &$+$2.19 &$+$0.79 &$+$35.5$\pm$0.1 & 400 & 20-11-2013 & 1$\times$1500 \\
          & 253 & 08 40 06.42 & $+$20 00 28.04 & 06.39 &$+$0.98 &$+$2.16 &$+$0.54 &$+$34.0$\pm$0.1 & 300 & 20-11-2013 & 1$\times$1200 \\
          & 283 & 08 40 22.09 & $+$19 40 11.78 & 06.39 &$+$1.02 &$+$2.20 &$+$0.60 &$+$34.7$\pm$0.1 & 390 & 20-11-2013 & 1$\times$1500 \\
 NGC 6633 & 100 & 18 27 54.73 & $+$06 36 00.33 & 08.31 &$+$1.13 &$+$2.64 &$+$0.66 &$-$29.1$\pm$0.2 & 350 & 15-07-2013 & 2$\times$1800 \\
          & 119 & 18 28 17.64 & $+$06 46 00.05 & 08.98 &$+$1.04 &$+$2.52 &$+$0.60 &$-$29.1$\pm$0.1 & 160 & 15-07-2013 & 1$\times$1800 \\
 NGC 6940 &  67 & 20 34 04.12 & $+$28 16 48.62 & 10.96 &$+$1.13 &$+$2.44 &$+$0.62 &$+$07.8$\pm$0.2 & 130 & 13-10-2013 & 2$\times$1800 \\
          &  69 & 20 34 05.75 & $+$28 11 18.38 & 11.63 &$+$1.13 &$+$2.69 &$+$0.62 &$+$07.9$\pm$0.2 & 160 & 13-10-2013 & 2$\times$1800 \\
          & 139 & 20 34 47.61 & $+$28 14 47.25 & 11.35 &$+$1.09 &$+$2.56 &$+$0.66 &$+$07.5$\pm$0.2 & 180 & 13-10-2013 & 1$\times$1800 \\
 NGC 7209 &  77 & 22 05 09.93 & $+$46 31 25.27 & 10.11 &$+$1.13 &$+$2.68 &$+$0.66 &$-$18.5$\pm$0.1 & 340 & 19-11-2013 & 2$\times$1800 \\
          &  89 & 22 05 17.62 & $+$46 29 00.64 & 09.45 &$+$1.35 &$+$3.64 &$+$0.90 &$-$18.8$\pm$0.1 & 250 & 19-11-2013 & 2$\times$1800 \\
          &  95$^{a}$ & 22 05 22.23 & $+$46 32 04.40 & 10.60 &$+$1.08 &$+$2.69 &$+$0.62 &$-$07.1$\pm$0.2 & 150 & 19-11-2013 & 2$\times$1800 \\

\hline
\end{tabular} 
\flushleft
Note: $^{a}$ Spectroscopic binary.
\end{table*}

\section{Observations and data reduction}
The sample of red giants in each of the target clusters was extracted from the {\small WEBDA}\footnote{\url{http://www.univie.ac.at/webda/}} database and the astrometric and photometric measurements are cross-checked with the {\small SIMBAD}\footnote{\url{http://simbad.u-strasbg.fr/simbad/}} astronomical database. 

High-resolution and high signal-to-noise (S/N) ratio optical spectra of 32 red giants across the twelve OCs were acquired during 2011 November and 2013 July, October, November and December with the Robert G. Tull coud\'{e} cross-dispersed echelle spectrograph (Tull et al. 1995) at the 2.7-m Harlan J. Smith reflector of the McDonald observatory. On all occasions we employed a 2048$\times$2048 24 $\mu$m pixel CCD detector and 52.67 grooves mm$^{-1}$ echelle grating with exposures centred at 5060 \AA. For each target, we obtained two to three exposures, each lasting for 20-30 min to minimize the influence of cosmic rays.

The spectra were reduced to 1D in multiple steps using various routines available within the \textit{imred} and \textit{echelle} packages of the standard spectral reduction software {\small IRAF}\footnote{IRAF is a general purpose software system for the reduction and analysis of astronomical data distributed by NOAO, which is operated by the Association of Universities for Research in Astronomy, Inc. under cooperative agreement with the National Science Foundation.} and then wavelength calibrated using Th-Ar spectra as a reference. The wavelength range 3600$-$9800 \AA\ spread across various echelle orders covered in a single exposure is sufficient to perform an abundance analysis of elements sampling all the major processes of stellar nucleosynthesis.

All but the spectra of giant stars in NGC 2287, NGC 6633 and NGC 7209 correspond to a resolving power of $R$ $\gtrsim$ 40,000 ($<$ 7.5 km s$^{-1}$) as measured by the FWHM of Th {\scs I} lines in comparison spectra, while the latter spectra were taken at a higher resolution of about 55,000. Multiple spectra were combined to acquire a single, high S/N spectra for each star. The combined spectra of each star has S/N ratios over 100 around 6000 \AA\ region, while at wavelengths shorter than 4000 \AA\ the S/N ratio decreases and falls to 15 around 3600 \AA\ region. We computed heliocentric the radial velocities (RVs) from the observed RV of each star using {\scs IRAF}'s {\it rvcorrect} routine. Our mean RV measurements for all OCs are in fair agreement with the previous estimates in the literature (Mermilliod et al. 2008). The methods of observations, data reduction and RV measurements are described in detail in Reddy et al. (2012, 2013, 2015). 

\begin{figure*}
\begin{center}
\includegraphics[trim=0.4cm 0.6cm 2.5cm 4.2cm, clip=true,width=0.95\textwidth,height=0.5\textheight]{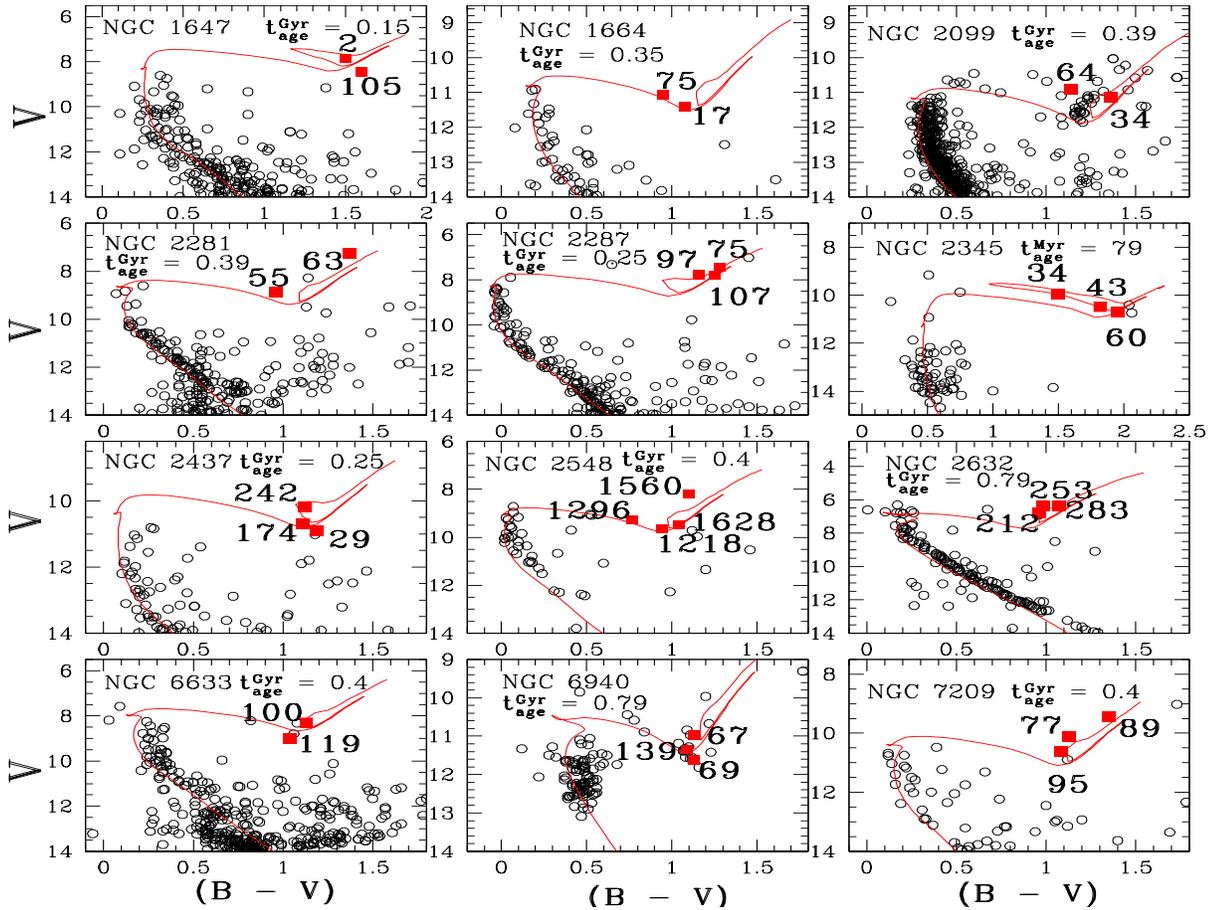}
\caption[]{The fitting of Padova isochrones (Marigo et al. 2008) constructed for [Fe/H]=$-$0.10 dex to the BV color-magnitude diagram of OCs in Table \ref{log_observations}. The photometric measurements of OCs are drawn from Francic (1989; NGC 1647, NGC 2281, NGC 6633), Hog et al. (1961; NGC 1664, NGC 7209), Kiss et al. (2001; NGC 2099), Sharma et al. (2006; NGC 2287), Moffat (1974; NGC 2345), Henden (2003; NGC 2437), Pesch (1961; NGC 2548), Johnson (1952; NGC 2632), and Walker (1958; NGC 6940) while the cluster's distance, age and reddening information is taken from the catalogue of Dias et al. (2002). The position of the program stars observed for abundance analysis is marked with red filled squares.} 
\label{cmd}
\end{center}
\end{figure*}

The membership of the stars to cluster was confirmed previously (see, {\scs WEBDA} database for details) through their proper motions, radial velocities and positions of stars in a photometric colour-magnitude diagram (see Figure \ref{cmd}). It is clear from the (B-V) colour-magnitude diagram in figure \ref{cmd} that the stars selected for abundance analysis from each of the OCs occupy either the red giant branch or red clump regions of the color-magnitude diagram. We have verified the membership of our program stars in each of the OCs using their radial velocities. Our mean RV measurements for all the cluster members are in fair agreement with the previous RV measurements for the red giants in OCs (Mermilliod et al. 2008).

The journal of observations for each of the cluster member is summarized in Table \ref{log_observations} together with the available optical and 2MASS\footnote{\url{http://irsa.ipac.caltech.edu/applications/Gator}} photometry (Cutri et al. 2003)\footnote{Originally published by the University of Massachusetts and Infrared Processing and Analysis Center (IPAC)/ California Institute of Technology.}, computed heliocentric RVs, and S/N ratios of the spectra around 6000 \AA. 

\begin{table*}
\centering
\begin{minipage}{150mm}
\caption{Basic photometric and spectroscopic atmospheric parameters for the stars in each cluster.}
\label{stellar_param}
\begin{tabular}{lcccccccccc}  \hline
\multicolumn{1}{l}{Cluster}& \multicolumn{1}{c}{Star ID} & \multicolumn{3}{c}{T$^{\rm phot}_{\rm eff}$ (K)} &
\multicolumn{1}{c}{$\log g^{(V-K)}_{\rm phot}$}& \multicolumn{1}{c}{T$^{\rm spec}_{\rm eff}$}& 
\multicolumn{1}{c}{$\log g^{\rm spec}$}& \multicolumn{1}{c}{$\xi^{\rm spec}_{t}$}&
\multicolumn{2}{c}{$\log(L/L_\odot)$} \\  
\cline{3-5}
\cline{10-11}
\multicolumn{1}{c}{}& \multicolumn{1}{c}{} & \multicolumn{1}{c}{(B-V)}& (V-K)& (J-K)& 
\multicolumn{1}{c}{(cm s$^{-2}$)}& \multicolumn{1}{c}{(K)} & \multicolumn{1}{c}{(cm s$^{-2}$)}& 
\multicolumn{1}{c}{(km s$^{-1}$)}& \multicolumn{1}{c}{spec} & \multicolumn{1}{c}{phot} \\
\hline

 NGC 1647 &    2 & 4568 & 4511 & 4516 & 1.42 & 4650 & 1.10 & 2.35 & 3.3 & 3.0  \\
          &  105 & 4410 & 4221 & 4140 & 1.65 & 4600 & 1.65 & 1.96 & 2.8 & 2.6  \\
 NGC 1664 &   17 & 5166 & 5117 & 4841 & 2.78 & 5150 & 2.90 & 1.79 & 1.7 & 1.8  \\
          &   75 & 5493 & 5417 & 4873 & 2.79 & 5200 & 2.90 & 1.71 & 1.7 & 1.9  \\
 NGC 2099 &   34 & 5138 & 5096 & 4866 & 2.38 & 5000 & 2.60 & 1.71 & 2.0 & 2.2  \\
          &   64 & 4692 & 4921 & 4648 & 2.34 & 5050 & 3.00 & 1.78 & 1.6 & 2.2  \\
 NGC 2281 &   55 & 5045 & 4971 & 4834 & 2.59 & 5200 & 3.10 & 1.70 & 1.6 & 2.0  \\
          &   63 & 4335 & 4284 & 4119 & 1.52 & 4400 & 1.65 & 1.78 & 2.7 & 2.8  \\
 NGC 2287 &   75 & 4384 & 4382 & 4181 & 1.51 & 4350 & 1.10 & 1.76 & 3.2 & 2.8  \\
          &   97 & 4572 & 4554 & 4388 & 1.77 & 4600 & 1.70 & 1.85 & 2.7 & 2.6  \\
          &  107 & 4588 & 4561 & 4498 & 1.78 & 4650 & 1.70 & 1.81 & 2.7 & 2.6  \\
 NGC 2345 &   34 & 5035 & 4552 & 4533 & 0.97 & 4850 & 0.85 & 3.23 & 3.7 & 3.4  \\
          &   43 & 4465 & 4306 & 4211 & 1.06 & 4300 & 1.20 & 2.21 & 3.1 & 3.3  \\
          &   60 & 4449 & 4325 & 4153 & 0.97 & 4300 & 1.60 & 2.21 & 2.7 & 3.4  \\
 NGC 2437 &   29 & 4744 & 4578 & 4417 & 2.30 & 4750 & 2.20 & 1.62 & 2.4 & 2.1  \\
          &  174 & 4897 & 4987 & 4931 & 2.36 & 4850 & 2.10 & 1.72 & 2.4 & 2.2  \\
          &  242 & 4878 & 4740 & 4607 & 2.11 & 4750 & 2.10 & 1.70 & 2.4 & 2.4  \\
 NGC 2548 & 1218 & 5003 & 4973 & 4952 & 2.66 & 5150 & 2.80 & 1.63 & 1.8 & 1.9  \\
          & 1296 & 5401 & 5050 & 4737 & 2.60 & 5200 & 3.20 & 1.55 & 2.0 & 2.0  \\
          & 1560 & 4694 & 4425 & 4319 & 1.81 & 4450 & 1.90 & 1.76 & 2.5 & 2.6  \\
          & 1628 & 4804 & 4741 & 4625 & 2.47 & 4800 & 2.55 & 1.49 & 2.0 & 2.0  \\
 NGC 2632 &  212 & 4936 & 4895 & 4135 & 2.65 & 4900 & 2.60 & 1.45 & 1.9 & 1.9  \\
          &  253 & 4896 & 4928 & 4851 & 2.58 & 4900 & 2.60 & 1.53 & 1.9 & 2.0  \\
          &  283 & 4819 & 4885 & 4650 & 2.55 & 4800 & 2.50 & 1.51 & 2.0 & 2.0  \\
 NGC 6633 &  100 & 4901 & 4921 & 4773 & 2.52 & 4950 & 2.55 & 1.44 & 2.0 & 2.0  \\
          &  119 & 5093 & 5057 & 4976 & 2.97 & 5200 & 3.05 & 1.41 & 1.6 & 1.7  \\
 NGC 6940 &   67 & 4901 & 5150 & 4913 & 2.84 & 4950 & 2.60 & 1.46 & 1.9 & 1.8  \\
          &   69 & 4901 & 4862 & 4884 & 3.01 & 5050 & 3.00 & 1.48 & 1.6 & 1.5  \\
          &  139 & 4984 & 5011 & 4759 & 2.97 & 5000 & 2.80 & 1.48 & 1.8 & 1.6  \\
 NGC 7209 &   77 & 4873 & 4829 & 4721 & 2.24 & 4850 & 2.20 & 1.69 & 2.3 & 2.3  \\
          &   89 & 4483 & 4105 & 4089 & 1.60 & 4200 & 1.50 & 1.85 & 2.8 & 2.6  \\
          &   95 & 4975 & 4826 & 4875 & 2.45 & 5000 & 2.80 & 1.76 & 1.8 & 2.1  \\

\hline
\end{tabular}
\end{minipage}
\end{table*}

\section{Abundance analysis} 
As the strength of a spectral line is influenced by the physical conditions in the stellar atmosphere and number density of absorbers, it is required to determine the stellar parameters to estimate the chemical abundances. Following the precepts discussed in Reddy et al. (2012), we derived the star's effective temperature, T$^{\rm phot}_{\rm eff_\star}$, by substituting the dereddened\footnote{The adopted interstellar extinctions are (A$_{V}$, A$_{K}$, E(V-K), E(J-K))= (3.1, 0.28, 2.75, 0.54)*E(B-V), where E(B-V) is taken from {\scs WEBDA}} optical and 2MASS photometric colors (B-V), (V-K$_{\rm s}$) and (J-K$_{\rm s}$) into the infrared flux method metallicity-dependent color$-$temperature calibrations of Alonso et al. (1999).

The surface gravities, log~$g$, were computed by incorporating the known distance to the OCs, T$^{\rm phot}_{\rm eff_\star}$, bolometric correction $BC_{V}$, cluster turn-off mass $M_{\star}$ and the solar values of T$_{\rm eff},_{\odot}$= 5777 K and log~{g}$_{\odot}$= 4.44 cm s$^{-2}$ into the well known log~$g$\,$-$\,T$_{\rm eff}$ relation (Reddy et al. 2012). Estimates of $BC_{V}$s were made using Alonso et al.'s (1999) relation connecting the photometric T$_{\rm eff_\star}$s and metallicities. 

We performed a differential abundance analysis relative to the Sun by running the {\it abfind} driver of {\scs \bf MOOG}\footnote{{\scs \bf MOOG} was developed and updated by Chris Sneden and originally described in Sneden (1973)} adopting the 1D model atmospheres and the iron line equivalent widths (EWs) following the local thermodynamic equilibrium (LTE) abundance analysis technique described in Reddy et al. (2015). The line EWs for a selected sample of clean, unblended, isolated and symmetric spectral lines of various atomic/ionic species were measured manually using routines in {\scs IRAF} while avoiding the wavelength regions affected by telluric contamination and heavy line crowding. We prepared a linelist of 300 absorption lines covering 23 elements from Na-Eu in the spectral range 4450$-$8850 \AA, as described in Reddy et al. (2015). The model atmospheres were interpolated linearly from the ATLAS9 model atmosphere grid of Castelli \& Kurucz (2003). 

The spectroscopic atmospheric parameters (T$_{\rm eff}$, log~$g$ and $\xi_{t}$) were derived by forcing the model-generated iron line EWs to match the observed ones by imposing the conditions of excitation and ionization equilibrium and the independence between the iron abundances and line's reduced EWs. We consider that the excitation equilibrium is satisfying whenever the slope between log\,$\epsilon$(Fe {\scs I}) and line's lower excitation potential (LEP) was $<$\,0.005 dex/$eV$ and that the ionization equilibrium is satisfied for $\mid$\,log\,$\epsilon$(Fe {\scs I})-log\,$\epsilon$(Fe {\scs II})\,$\mid$$\leq$\,0.02 dex. The adopted $\xi_{t}$ is considered satisfactory when the slope between the Fe abundance from Fe {\scs II} lines and the line's reduced EWs was almost zero. As all these stellar parameters are interdependent, several iterations are required to extract a suitable model from the grid of stellar atmospheres so that all the spectral lines in observed spectra are readily reproduced. The conditions of the excitation and ionization equilibrium and a check on the $\xi_{t}$ measured from iron lines have been verified with a suite of 
Ni, Sc, Ti, V and Cr lines, as described in detail in Reddy et al. (2015). Following Reddy et al. (2015), typical errors are $\pm$75 K in $T_{\rm eff}$, 0.25 cm s$^{-2}$ in log~$g$ and 0.20 km s$^{-1}$ in $\xi_{t}$. The stellar parameters estimated for program stars are given in Table \ref{stellar_param}.  

The abundance analysis was extended to other species in the linelist using EWs but synthetic profiles were computed for lines affected by hyperfine structure (hfs) and isotopic splitting and/or affected by blends. Our linelists have been tested extensively to reproduce the solar and Arcturus spectra before applying them to selected spectral features in the stellar spectra of program stars. We used a standard synthetic profile fitting procedure by running the {\it synth} driver of {\scs \bf MOOG} adopting the spectroscopically determined stellar parameters. 

\begin{table*}
{\fontsize{8}{8}\selectfont
\caption[Elemental abundances for stars in the OCs NGC 1647 \& 1664]{The chemical abundances of red giants in OCs NGC 1647 and NGC 1664. The abundances measured by synthesis are presented in bold typeface while the remaining elemental abundances were calculated using the line EWs. Numbers in the parentheses indicate the number of lines used in calculating the abundance of that element. } 
\vspace{0.2cm}
\label{abu_NGC1647}
\begin{tabular}{ccccccc}   \hline
\multicolumn{1}{l}{Species} & \multicolumn{1}{c}{NGC 1647$\#$2} & \multicolumn{1}{c}{NGC 1647$\#$105} &\multicolumn{1}{c}{NGC 1647$_{\mbox{Avg.}}$} \vline & \multicolumn{1}{c}{NGC 1664$\#$17} & \multicolumn{1}{c}{NGC 1664$\#$75} & \multicolumn{1}{c}{NGC 1664$_{\mbox{Avg.}}$}  \\ \hline
 
$[$Na I/Fe$]$  &$+0.43\pm0.04$(4) &$+0.31\pm0.05$(3)  &$+0.37\pm0.05$ &$+0.27\pm0.03$(5) &$+0.28\pm0.03$(6) &$+0.27\pm0.02$   \\
$[$Mg I/Fe$]$  &$+0.26\pm0.03$(2) &$+0.16\pm0.05$(2)  &$+0.21\pm0.04$ &$-0.04\pm0.03$(3) &$-0.08\pm0.04$(4) &$-0.06\pm0.03$   \\
$[$Al I/Fe$]$  &$+0.10\pm0.05$(5) &$+0.16\pm0.04$(4)  &$+0.13\pm0.05$ &$\,0.00\pm0.01$(5) &$-0.03\pm0.03$(5) &$-0.01\pm0.02$  \\
 $\vdots$        &   $\vdots$       &   $\vdots$         &   $\vdots$     &   $\vdots$        &   $\vdots$        &   $\vdots$    \\
$[$Eu II/Fe$]$ & $\bf+0.13$(1)    & $\bf+0.10$(1)      &$\bf+0.11$     & $\bf+0.22$(1)     & $\bf+0.20$(1)     &$\bf+0.21$     \\
 
\hline
\end{tabular} }
\flushleft{ {\bf Note} -- Only a portion of this table is shown here for guidance regarding its form and content. The full table is available with the online version of the paper.}
\end{table*}

The chemical abundances for the individual cluster members averaged over all available lines of given species are presented in Tables 3$-$12\footnote{Tables 3$-$12 are available with the online version of the paper.}, relative to solar abundances derived from the adopted $gf$-values (see Table 4 from Reddy et al. 2012). Table entries provide the average [Fe/H] and [X/Fe] for all elements, and standard deviation along with the number of lines used in calculating the abundance of that element. Within the run from Na to Eu, all stars in a given cluster have very similar [X/Fe] for almost all the elements. 
Following Reddy et al. (2015), we evaluated the impact of stellar parameters (T$_{\rm eff}$, log~$g$, $\xi_{t}$) on the derived abundances and the abundance differences caused by the variation of each stellar parameter with respect to the best model parameter are summed in quadrature to obtain a global uncertainty $\sigma_{2}$. The total error $\sigma_{tot}$ for each of the element is the quadratic sum of $\sigma_{1}$ and $\sigma_{2}$. The final chemical content of each OC along with the $\sigma_{tot}$ from this study is presented in Table 13.

\begin{landscape}
\begin{table}
{\fontsize{5}{7}\selectfont
\begin{tabular}{lcccccccccccc}   \hline
\multicolumn{1}{c}{Species}  & \multicolumn{1}{c}{NGC 1647} & \multicolumn{1}{c}{NGC 1664} & \multicolumn{1}{c}{NGC 2099} & \multicolumn{1}{c}{NGC 2281} & \multicolumn{1}{c}{NGC 2287} & \multicolumn{1}{c}{NGC 2345} & \multicolumn{1}{c}{NGC 2437} & \multicolumn{1}{c}{NGC 2548} & \multicolumn{1}{c}{NGC 2632} & \multicolumn{1}{c}{NGC 6633} & \multicolumn{1}{c}{NGC 6940} & \multicolumn{1}{c}{NGC 7209}  \\ 
\multicolumn{1}{l}{Age (Myr)}  & \multicolumn{1}{c}{$=>$ 144} & \multicolumn{1}{c}{292} & \multicolumn{1}{c}{347} & \multicolumn{1}{c}{358} & \multicolumn{1}{c}{251} & \multicolumn{1}{c}{71} & \multicolumn{1}{c}{251} & \multicolumn{1}{c}{398} & \multicolumn{1}{c}{729} & \multicolumn{1}{c}{425} & \multicolumn{1}{c}{721} & \multicolumn{1}{c}{414}  \\ \hline

$[$Na I/Fe$]$ &$+0.37\pm0.05$ &$+0.27\pm0.02$ &$+0.23\pm0.02$ &$+0.20\pm0.03$ &$+0.24\pm0.03$ &$+0.18\pm0.02$ &$+0.32\pm0.02$ &$+0.26\pm0.02$ &$+0.35\pm0.03$ &$+0.20\pm0.02$ &$+0.25\pm0.03$ &$+0.27\pm0.02$ \\
$[$Mg I/Fe$]$ &$+0.21\pm0.04$ &$-0.06\pm0.03$ &$-0.07\pm0.03$ &$-0.05\pm0.03$ &$+0.10\pm0.03$ &$+0.05\pm0.03$ &$-0.03\pm0.02$ &$-0.01\pm0.02$ &$-0.02\pm0.03$ &$-0.05\pm0.03$ &$-0.04\pm0.03$ &$+0.01\pm0.02$ \\
$[$Al I/Fe$]$ &$+0.13\pm0.05$ &$-0.01\pm0.02$ &$+0.01\pm0.03$ &$-0.02\pm0.02$ &$+0.05\pm0.02$ &$+0.03\pm0.02$ &$-0.01\pm0.02$ &$+0.03\pm0.02$ &$+0.05\pm0.02$ &$+0.06\pm0.02$ &$+0.05\pm0.03$ &$+0.06\pm0.02$ \\
$[$Si I/Fe$]$ &$+0.21\pm0.05$ &$+0.10\pm0.03$ &$+0.10\pm0.03$ &$+0.18\pm0.03$ &$+0.22\pm0.03$ &$+0.25\pm0.02$ &$+0.13\pm0.03$ &$+0.11\pm0.02$ &$+0.16\pm0.03$ &$+0.06\pm0.03$ &$+0.14\pm0.03$ &$+0.22\pm0.03$ \\
$[$Ca I/Fe$]$ &$-0.02\pm0.04$ &$-0.03\pm0.03$ &$-0.06\pm0.04$ &$+0.11\pm0.03$ &$-0.05\pm0.03$ &$-0.13\pm0.03$ &$-0.02\pm0.03$ &$-0.01\pm0.02$ &$-0.01\pm0.03$ &$\,0.00\pm0.03$ &$\,0.00\pm0.03$ &$-0.05\pm0.02$ \\
$[$Sc II/Fe$]$&$\bf\,0.00\pm0.05$ &$\bf\,0.00\pm0.05$ &$\bf-0.06\pm0.05$ &$\bf+0.02\pm0.02$ &$\bf-0.05\pm0.05$ &$\bf+0.06\pm0.05$ &$\bf+0.04\pm0.05$ &$\bf+0.01\pm0.05$ &$\bf-0.02\pm0.05$ &$\bf-0.05\pm0.05$ &$\bf-0.04\pm0.05$ &$\bf-0.04\pm0.05$ \\
$[$Ti I/Fe$]$ &$-0.15\pm0.03$ &$+0.06\pm0.03$ &$+0.04\pm0.03$ &$-0.01\pm0.03$ &$-0.14\pm0.03$ &$-0.05\pm0.02$ &$-0.13\pm0.03$ &$-0.01\pm0.03$ &$-0.08\pm0.03$ &$-0.04\pm0.03$ &$-0.04\pm0.02$ &$-0.08\pm0.02$ \\
$[$Ti II/Fe$]$&$-0.14\pm0.03$ &$+0.03\pm0.03$ &$\,0.00\pm0.03$ &$-0.04\pm0.03$ &$-0.11\pm0.03$ &$-0.03\pm0.02$ &$-0.06\pm0.02$ &$-0.03\pm0.02$ &$-0.09\pm0.03$ &$-0.06\pm0.03$ &$-0.09\pm0.02$ &$-0.13\pm0.02$ \\
$[$V I/Fe$]$ &$-0.11\pm0.05$ &$+0.08\pm0.03$ &$+0.13\pm0.03$ &$+0.06\pm0.03$ &$-0.07\pm0.03$ &$-0.02\pm0.03$ &$-0.10\pm0.03$ &$+0.08\pm0.03$ &$+0.03\pm0.03$ &$-0.02\pm0.03$ &$+0.07\pm0.03$ &$+0.04\pm0.03$ \\
$[$Cr I/Fe$]$ &$-0.06\pm0.04$ &$+0.05\pm0.02$ &$+0.04\pm0.03$ &$\,0.00\pm0.02$ &$-0.03\pm0.03$ &$-0.01\pm0.02$ &$-0.03\pm0.02$ &$+0.02\pm0.02$ &$+0.02\pm0.03$ &$-0.01\pm0.03$ &$+0.03\pm0.03$ &$\,0.00\pm0.03$ \\
$[$Cr II/Fe$]$&$+0.07\pm0.03$ &$+0.08\pm0.03$ &$+0.09\pm0.02$ &$+0.06\pm0.02$ &$+0.09\pm0.02$ &$+0.06\pm0.02$ &$+0.08\pm0.02$ &$+0.06\pm0.03$ &$+0.09\pm0.03$ &$+0.04\pm0.02$ &$+0.03\pm0.03$ &$+0.13\pm0.02$ \\
$[$Mn I/Fe$]$ &$\bf-0.26\pm0.05$ &$\bf-0.09\pm0.05$ &$\bf-0.08\pm0.05$ &$\bf-0.04\pm0.05$ &$\bf-0.20\pm0.05$ &$\bf-0.21\pm0.05$ &$\bf+0.09\pm0.05$ &$\bf-0.08\pm0.05$ &$\bf-0.06\pm0.05$ &$\bf-0.05\pm0.05$ &$\bf-0.05\pm0.05$ &$\bf-0.09\pm0.05$ \\
$[$Fe I/H$]$ &$-0.15\pm0.05$ &$-0.11\pm0.04$ &$-0.01\pm0.03$ &$-0.05\pm0.04$ &$-0.19\pm0.03$ &$-0.26\pm0.03$ &$-0.18\pm0.03$ &$-0.11\pm0.03$ &$+0.08\pm0.03$ &$-0.05\pm0.03$ &$-0.06\pm0.03$ &$-0.08\pm0.03$ \\
$[$Fe II/H$]$&$-0.14\pm0.05$ &$-0.10\pm0.03$ &$-0.02\pm0.03$ &$-0.05\pm0.03$ &$-0.18\pm0.03$ &$-0.26\pm0.03$ &$-0.17\pm0.03$ &$-0.11\pm0.02$ &$+0.08\pm0.03$ &$-0.04\pm0.03$ &$-0.07\pm0.03$ &$-0.05\pm0.03$ \\
$[$Co I/Fe$]$ &$-0.08\pm0.05$ &$+0.07\pm0.03$ &$+0.05\pm0.02$ &$+0.05\pm0.02$ &$+0.05\pm0.02$ &$-0.01\pm0.02$ &$+0.05\pm0.03$ &$+0.03\pm0.03$ &$+0.08\pm0.03$ &$-0.01\pm0.03$ &$+0.06\pm0.03$ &$+0.07\pm0.02$ \\
$[$Ni I/Fe$]$ &$-0.08\pm0.05$ &$-0.03\pm0.03$ &$\,0.00\pm0.03$ &$-0.04\pm0.03$ &$-0.03\pm0.03$ &$-0.07\pm0.02$ &$-0.03\pm0.02$ &$-0.02\pm0.02$ &$+0.03\pm0.03$ &$-0.04\pm0.03$ &$\,0.00\pm0.03$ &$-0.02\pm0.03$ \\
$[$Cu I/Fe$]$ &$\bf-0.21\pm0.05$ &$\bf-0.19\pm0.05$ &$\bf-0.10\pm0.05$ &$\bf-0.09\pm0.05$ &$\bf-0.17\pm0.05$ &$\bf-0.28\pm0.05$ &$\bf-0.05\pm0.05$ &$\bf-0.14\pm0.05$ &$\bf-0.08\pm0.05$ &$\bf-0.15\pm0.05$ &$\bf-0.14\pm0.05$ &$\bf-0.14\pm0.05$ \\
$[$Zn I/Fe$]$ &$\bf-0.19\pm0.05$ &$\bf-0.17\pm0.05$ &$\bf-0.07\pm0.05$ &$\bf-0.09\pm0.05$ &$\bf-0.17\pm0.05$ &$\bf-0.22\pm0.05$ &$\bf-0.03\pm0.05$ &$\bf-0.13\pm0.05$ &$\bf-0.07\pm0.05$ &$\bf-0.13\pm0.05$ &$\bf-0.03\pm0.05$ &$\bf-0.15\pm0.05$ \\
$[$Y II/Fe$]$&$+0.09\pm0.05$ &$+0.10\pm0.03$ &$+0.10\pm0.03$ &$+0.13\pm0.03$ &$+0.01\pm0.02$ &$+0.16\pm0.02$ &$+0.08\pm0.02$ &$+0.06\pm0.03$ &$-0.05\pm0.03$ &$+0.04\pm0.03$ &$+0.05\pm0.03$ &$+0.06\pm0.03$ \\
$[$Zr I/Fe$]$ &$-0.05\pm0.03$ &$+0.18\pm0.00$ &$+0.18\pm0.02$ &$+0.08\pm0.03$ &$-0.06\pm0.02$ &$-0.05\pm0.03$ &$+0.08\pm0.02$ &$+0.08\pm0.02$ &$+0.03\pm0.03$ &$+0.09\pm0.02$ &$+0.08\pm0.02$ &$+0.02\pm0.03$ \\
$[$Ba II/Fe$]$&$\bf+0.51\pm0.05$ &$\bf+0.13\pm0.05$ &$\bf+0.09\pm0.05$ &$\bf+0.18\pm0.05$ &$\bf+0.27\pm0.05$ &$\bf+0.27\pm0.05$ &$\bf+0.37\pm0.05$ &$\bf+0.11\pm0.05$ &$\bf-0.08\pm0.05$ &$\bf+0.17\pm0.05$ &$\bf+0.10\pm0.05$ &$\bf+0.06\pm0.05$ \\
$[$La II/Fe$]$&$-0.17\pm0.04$ &$+0.15\pm0.03$ &$+0.03\pm0.03$ &$+0.02\pm0.03$ &$-0.09\pm0.02$ &$+0.01\pm0.02$ &$-0.07\pm0.02$ &$-0.04\pm0.03$ &$-0.06\pm0.03$ &$\,0.00\pm0.03$ &$-0.08\pm0.02$ &$-0.09\pm0.02$ \\
$[$Ce II/Fe$]$&$+0.10\pm0.02$ &$+0.27\pm0.02$ &$+0.11\pm0.03$ &$+0.08\pm0.02$ &$+0.02\pm0.02$ &$+0.02\pm0.02$ &$+0.09\pm0.02$ &$+0.11\pm0.02$ &$-0.13\pm0.02$ &$+0.06\pm0.02$ &$+0.01\pm0.02$ &$+0.03\pm0.03$ \\
$[$Nd II/Fe$]$&$+0.02\pm0.03$ &$+0.21\pm0.03$ &$+0.08\pm0.05$ &$+0.16\pm0.02$ &$+0.07\pm0.02$ &$+0.03\pm0.02$ &$+0.10\pm0.02$ &$+0.08\pm0.03$ &$\,0.00\pm0.03$ &$+0.12\pm0.02$ &$+0.07\pm0.03$ &$+0.10\pm0.03$ \\
$[$Sm II/Fe$]$&$-0.04\pm0.04$ &$+0.22\pm0.03$ &$+0.07\pm0.03$ &$+0.19\pm0.03$ &$+0.09\pm0.02$ &$+0.12\pm0.02$ &$+0.02\pm0.02$ &$+0.07\pm0.02$ &$+0.06\pm0.03$ &$+0.12\pm0.03$ &$+0.05\pm0.03$ &$+0.07\pm0.02$ \\
$[$Eu II/Fe$]$&$\bf+0.11\pm0.05$ &$\bf+0.21\pm0.05$ &$\bf+0.01\pm0.05$ &$\bf+0.06\pm0.05$ &$\bf+0.08\pm0.05$ &$\bf+0.12\pm0.05$ &$\bf+0.09\pm0.05$ &$\bf+0.10\pm0.05$ &$\bf+0.01\pm0.05$ &$\bf+0.12\pm0.05$ &$\bf+0.04\pm0.05$ &$\bf+0.06\pm0.05$ \\

\hline
\end{tabular}
 } 
\flushleft{{\bf Table 13.} Mean abundance ratios, [X/Fe], for elements from Na to Eu for the OCs NGC 1647, 1664, 2099, 2281, 2287, 2345, 2437, 2548, 2632, 6633, 6940 and 7209 from this study. Abundances calculated by synthesis are presented in bold typeface. The cluster ages are taken from the catalogue of Dias et al. (2002). }
\label{mean_abundance}
\end{table}
 \end{landscape}

\begin{table*} 
\flushleft{{\bf Table 14.} The input data used in the computation of Galactic orbits of 79 OCs in this study.}

\centering
\label{input_propermotion}
{\fontsize{7}{8}\selectfont
\begin{tabular}{lllcrccccr@{}}  \hline
\multicolumn{1}{l}{Cluster}& \multicolumn{1}{c}{$\alpha$(2000.0)}& \multicolumn{1}{r}{$\delta$(2000.0)}& \multicolumn{1}{c}{l}& 
\multicolumn{1}{c}{b}& \multicolumn{1}{c}{d$_{\odot}$}& \multicolumn{1}{c}{RV} & \multicolumn{1}{c}{$\mu_{\alpha}$ cos $\delta$}& 
\multicolumn{1}{c}{$\mu_{\delta}$}& \multicolumn{1}{c}{proper motion} \\
\multicolumn{1}{l}{ }& \multicolumn{1}{c}{hh:mm:ss} & \multicolumn{1}{c}{$^{\circ}$ $^{'}$ $^{''}$}& \multicolumn{1}{c}{(deg.)} &
\multicolumn{1}{c}{(deg.)}& \multicolumn{1}{c}{(kpc)}& \multicolumn{1}{c}{(km sec$^{-1}$)}& \multicolumn{1}{r}{(mas yr$^{-1}$)}& 
\multicolumn{1}{r}{(mas yr$^{-1}$)}& {Reference}  \\
\hline 

 NGC  752 & 01:57:41 & +37:47:06 & 137.125 &$-$23.254 & 457 &   6.30$\pm$0.10 &  7.50$\pm$0.32 &$-$11.50$\pm$0.32 & TYC  \\
 NGC 1342 & 03:31:38 & +37:22:36 & 154.952 &$-$15.342 & 665 & $-$10.67$\pm$0.11 & $-$1.15$\pm$0.87 & $-$2.80$\pm$0.87 & D14\\
 NGC 1647 & 04:45:55 & +19:06:54 & 180.337 &$-$16.772 & 540 & $-$7.02$\pm$0.22  & $-$1.77$\pm$0.29 & $-$2.00$\pm$0.29 & D14\\
\hline

\end{tabular} }
\flushleft{ {\bf Sources} -- TYC: Dias et al. (2001, 2002); D14: Dias et al. (2014); K13: Kharchenko et al. (2013); BDW: Baumgardt et al. (2000); L03: Loktin \& Beshenov (2003); SOC: Soubiran et al. (2000); MRZ: Magrini et al. (2010); W09: Wu et al. (2009). \\
 {\bf Note} -- Only a portion of this table is shown here for guidance regarding its form and content. The full table is available with the online version of the paper.}
\end{table*}

\section{Results}
We merged our sample of twenty-eight OCs (12 from this study and 16 from our previous papers) with the available high-quality results in the literature to enlarge the dataset. Systematic offsets between chemical abundances of OCs collected from different resources and our sample of 28 OCs are negligible as we have remeasured the chemical content of all 51 OCs drawn from the literature using their published EWs, our models, linelists and reference solar abundances. The chemical abundances of 51 OCs are presented previously in Table 13 from Reddy et al. (2015). Stars in our sample and those drawn from the literature are similar red giants analysed identically and, thus, systematic errors affecting the abundances and abundance ratios [X/Fe] should be consistent (and small) across the sample which spans a narrow range in metallicity of $\sim$ $-$0.5 to 0.3 dex but a good range in ages and Galactocentric distances. Following the kinematic criteria described in Reddy et al. (2015), we assigned OCs either to the thin disc, thick disc or halo populations using the space velocity components of the cluster calculated using the proper motions, RVs, and heliocentric distances drawn mostly from the catalogue of Dias et al. (2002) and occasionally from the MWSC catalogue of Kharchenko et al. (2013) and literature resources as listed in the Table 14. We use the present sample of 79 OCs to look at abundance trends with cluster's age and Galactocentric distance. 

\subsection{Age-metallicity relation}
\begin{figure}
\begin{center}
\includegraphics[trim=1.7cm 0.2cm 0.5cm 0.8cm, clip=true,height=0.35\textheight,angle=-90]{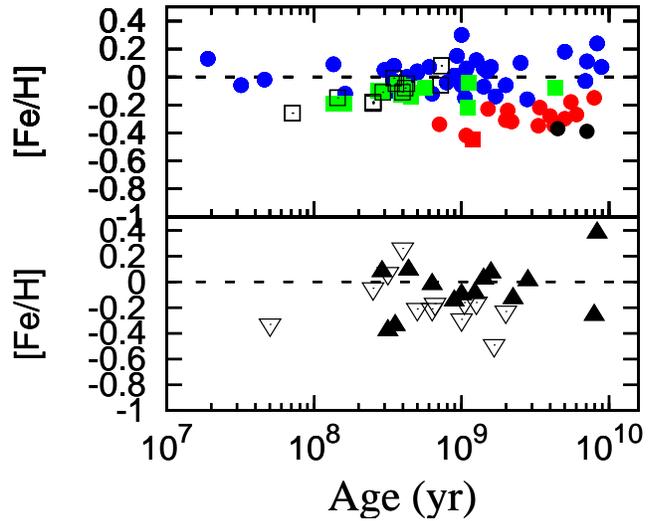}
\caption[]{ The distribution of iron abundance as a function of cluster's age. Top panel: The sample of 12 OCs analysed in this paper are represented as black open squares (thin disc) while our previous sample of 16 OCs are shown as green filled squares (thin disc) and one red filled square (thick disc). Clusters from the literature are presented as blue dots (thin disc), red dots (thick disc) and black dots (halo). Bottom panel: OCs from Frinchaboy et al. (2013) are designated with filled triangles (R$_{\rm gc}<$10 kpc) and inverted open triangles (10 $<$R$_{\rm gc}<$15 kpc). }
\label{fe_ageoc} 
\end{center}
\end{figure}

Figure \ref{fe_ageoc} shows the run of [Fe/H] with age for the full sample of 79 OCs covering ages from 100 Myr to 9 Gyr with a distinction made between thin, thick disc and halo members. The age distributions for these components are consistent with distributions for local field stars with the reasonable caveat that the age distribution of OCs is depleted at great ages because clusters are dissolved by internal dynamical effects and Galactic tidal forces. Dissolution times are short for small and poorly populated OCs but OCs with intermediate masses ($\sim$ 500 $M_\odot$ to 1000 $M_\odot$) can survive several Gyr, if they are located in the external regions of the disc (Pavani \& Bica 2007). A cluster of even $10,000\,M_{\odot}$ may dissolve in less than 2 Gyr in the solar vicinity from the frequent encounters with dense molecular clouds (Lamers \& Gieles 2006). Since disruption timescales must be much shorter well inside the solar circle, the survival of old OCs in Figure \ref{fe_ageoc} suggest that either such OCs were very massive or located exclusively in external (i.e., low density) regions of the Galaxy.

Metallicity$-$age relations for thin and thick disc field stars have been extensively studied -- see, for example, Bensby et al. (2005, 2014), Reddy et al. (2006). Thin disc field stars span the interval [Fe/H] from about $-0.8$ to $+0.4$ dex with little to no age dependence. In contrast, the thick disc is concentrated at sub-solar [Fe/H] (say $-0.2$ to $-1.0$) and is composed almost exclusively of stars with ages greater than about 8 Gyr with a steep trend of decreasing [Fe/H] with increasing age.

The sample of thin disc OCs matches that expected from thin disc field stars with the supplementary condition that the youngest clusters have yet to contribute young field stars. These OCs span the [Fe/H] range from about $-0.2$ to $+0.2$ with ages up to almost 10 Gyr. The thick disc OCs have a [Fe/H] about $-0.3$ dex lower than the thin disc OCs with an average age greater than the thin disc OCs.

Inspection of the [Fe/H] versus age suggests that young and old OCs with thin disc kinematics have very similar metallicities and display a flat trend with increasing age with bounds $\pm$0.2 dex for almost all OCs as compared with $\sigma$ for each cluster of about $\pm0.05$ dex or less. Not surprisingly, we notice a similar behaviour but with a smaller spread of about $\pm$0.1 dex for other iron-peak elements Sc, V, Cr, Co and Ni over the age range sampled by these thin disc OCs (Reddy et al. 2015). Since Fe and iron-peak elements are primarily products of Type Ia supernovae from white dwarfs which exceed the Chandrasekhar mass limit, these clusters may formed well after Type Ia supernovae had begun to make their contribution to the Galactic inventory. As the sampled thin disc OCs have a considerable spread in R$_{\rm gc}$ ($\sim$\,5 to 12 kpc), it is likely that the observed age$-$metallicity relation carries also the signature of metallicity evolution at different R$_{\rm gc}$s as the Galaxy evolved which probably represent the spread in [Fe/H] at any age over the lifetime of the disc.

The five youngest clusters with ages from 10 to 100 Myr serve as an illustrative example of spread in [Fe/H] caused by the metallicity enrichment at different R$_{\rm gc}$s in the Galaxy; OCs in the solar vicinity (blue dots), namely IC 2391 ([Fe/H]=$-0.01\pm0.05$ dex, 8.0 kpc, 46 Myr), IC 2602 ([Fe/H]=$-0.05\pm0.04$ dex, 7.9 kpc, 32 Myr), NGC 7160 ([Fe/H]=$+0.13\pm0.06$ dex, 8.2 kpc, 19 Myr) are metal-rich than clusters far away from the Sun, namely NGC 2345 (black open square, [Fe/H]=$-0.26\pm0.03$ dex, 9.7 kpc, 71 Myr; this study), NGC 2234 (inverted triangle, [Fe/H]=$-0.33\pm0.06$ dex, 12.7 kpc, 50 Myr; Frinchaboy et al. 2013). These clusters because of young ages may not have moved far away from their birthplaces and, thus, represent the chemical composition of gas in those locations in the Galactic disc which is compatible with the widely held impression of decreasing [Fe/H] with increasing R$_{\rm gc}$ (Friel et al. 2002, 2010; Magrini et al. 2010).

\begin{figure}
\begin{center}
\includegraphics[trim=1.7cm 0.2cm 0.5cm 0.8cm, clip=true,height=0.35\textheight,angle=-90]{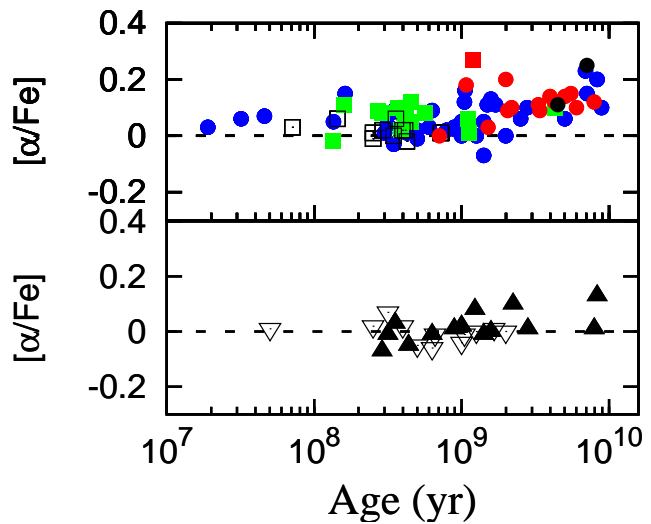}
\caption[]{Same as Figure \ref{fe_ageoc}, but for [$\alpha$/Fe] versus cluster's age.}
\label{alpha_ageoc} 
\end{center}
\end{figure}

Frinchaboy et al. (2013) have presented chemical abundances of 28 OCs derived from the automated processing of infrared spectra of cluster giants collected from the APOGEE survey in the tenth data release of the Sloan Digital Sky Survey (SDSS-III, DR10; Ahn et al. 2014). Of these 28 clusters, 15 clusters have measured chemical abundances from a single star whose membership to the cluster has been assumed based on its presence in the Dias et al. (2002) catalog based isochrone fit and a radial velocity membership probability of above 50\%. Here, we avoid assigning OCs from Frinchaboy et al. (2013) to Galactic populations, as almost all those clusters have no reported radial velocities even in the catalogue of Dias et al. (2002). We designate in figure \ref{fe_ageoc} and figure \ref{alpha_ageoc}, the chemical abundances of Frinchaboy et al. sample of clusters as filled triangles (R$_{\rm gc}<$10 kpc) and inverted open triangles (10 $<$R$_{\rm gc}<$15 kpc). 

\begin{figure*}
\begin{center}
\includegraphics[trim=0.1cm 5.9cm 3.2cm 4.65cm, clip=true,height=0.28\textheight,width=0.90\textwidth]{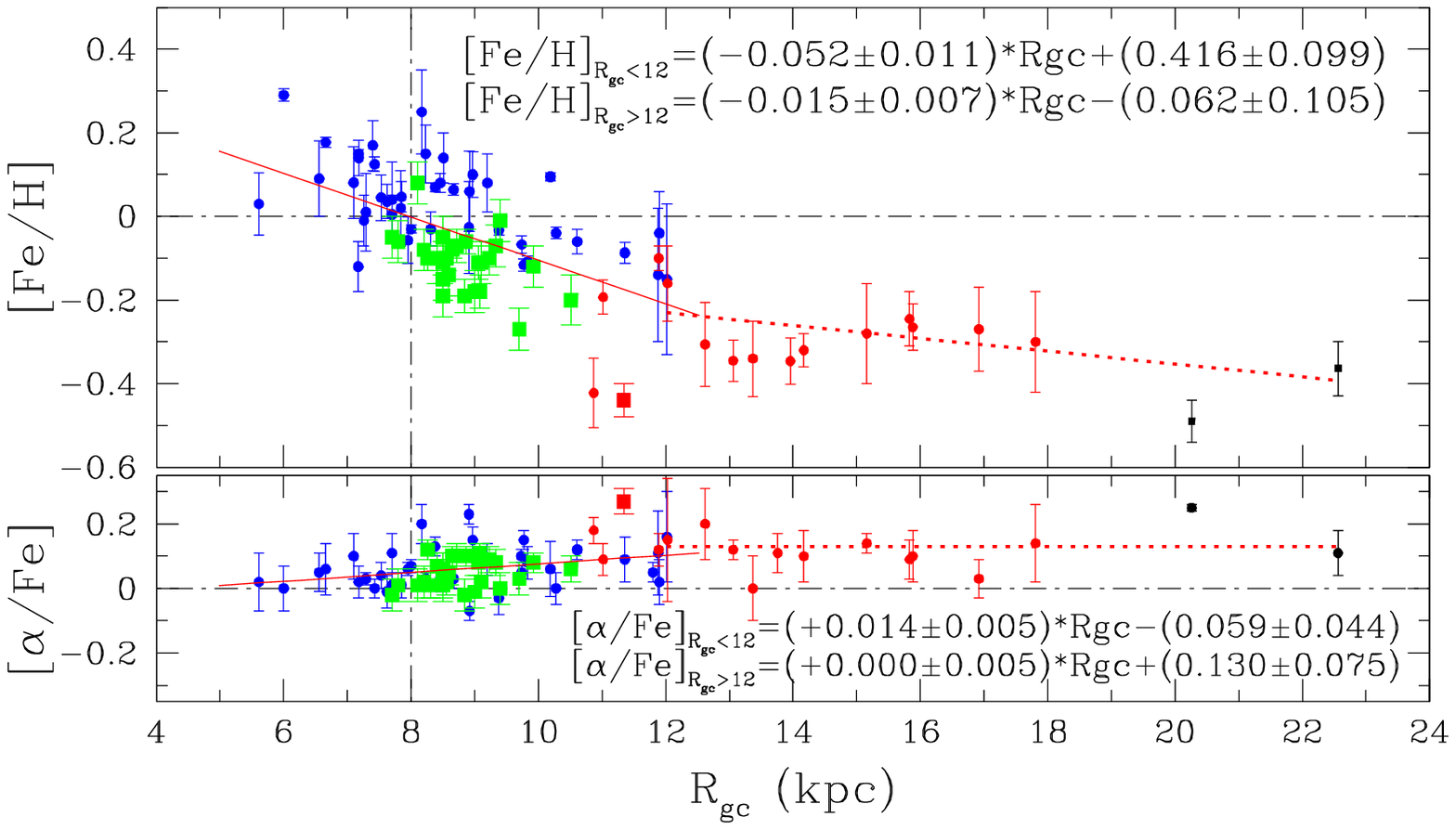}
\caption[]{The distribution of iron abundance as a function of cluster's distance from the Galactic center. Our sample of 28 OCs are represented as green filled squares (thin disc) and one red filled square (thick disc) while the literature sample is represented by filled black, blue and red dots as in figure \ref{fe_ageoc}. The horizontal and vertical dash-dotted lines represent, respectively, the solar mix of elements and the Galactocentric distance of the Sun (R$_{\rm gc}$=\,8.0 kpc).}
\label{gradient}
\end{center}
\end{figure*}

Though metallicities of OCs from the infrared study are independent of cluster's age, there appears to be an offset of about 0.15 dex between the distant OCs (inverted triangles) and the thick disc clusters (red dots) which as we show in figure \ref{gradient} are populating the Galactic disc in the radial range 11 to 24 kpc. Note however that systematic offsets are typical of what is seen among similar abundance analyses by different authors of the same or similar stars, so the abundance differences found here is not surprising.

\subsection{[$\alpha$/Fe] vs. cluster's age}
The run of [$\alpha$/Fe] with age where the $\alpha$-elements comprise Mg, Si, Ca and Ti is shown Figure \ref{alpha_ageoc}. We notice a weak correlation between [$\alpha$/Fe] and age whose significance was obscured in comparisons of inhomogeneous data sets of the previous analyses of OCs (Bragaglia et al. 2008, Friel et al. 2010, Yong et al. 2012). There is an indication that almost all the OCs older than 1 Gyr yield slightly more positive [$\alpha$/Fe] ratios for both thin and thick disc samples than those represented by younger ages. A similar behaviour is observable for Galactic field stars where the older thin and thick disc field dwarfs from Bensby et al. (2014) yield more positive [$\alpha$/Fe] ratios than the younger field dwarfs with thin disc kinematics. The similarity in [$\alpha$/Fe] between the thin and thick disc clusters with ages of 4 to 9 Gyr suggest that age cannot be the key variable, but different proportions of Type II to Type Ia supernovae products in clusters formed at different locations in the Galaxy are in the main responsible. There is an indication that almost all the OCs older than 1 Gyr yield slightly more positive [$\alpha$/Fe] ratios for both thin and thick disc samples than those represented by younger ages.
A similar trend of decreasing [$\alpha$/Fe] from old to young clusters is visible in infrared data from Frinchaboy et al. (2013) but with an offset.

\begin{table*}
\flushleft{{\bf Table 15.} Current positions and Galactic space velocity components for the total sample of 79 OCs. The full table is available with the online version of the paper.}

\centering
\label{xyzuvw} 
{\fontsize{8}{7}\selectfont
\begin{tabular}{lcccrrrr@{}} \hline
\multicolumn{1}{l}{Cluster}& \multicolumn{1}{c}{x}& \multicolumn{1}{c}{y}& \multicolumn{1}{c}{z}& \multicolumn{1}{c}{U$_{\rm GSR}$}& \multicolumn{1}{c}{V$_{\rm GSR}$} & \multicolumn{1}{c}{W$_{\rm GSR}$}& \multicolumn{1}{c}{ Age }   \\ 
\multicolumn{1}{l}{ }& \multicolumn{1}{c}{(kpc)} & \multicolumn{1}{c}{(kpc)} & \multicolumn{1}{c}{(kpc)}& \multicolumn{1}{c}{(km s$^{-1}$)}& \multicolumn{1}{c}{(km s$^{-1}$)}& \multicolumn{1}{c}{(km s$^{-1}$)} & \multicolumn{1}{c}{(Myr)}  \\   \hline

\multicolumn{8}{c}{\bf \normalsize {Thin disk}} \\
 NGC  752 &  -8.308$\pm$0.031 &   0.285$\pm$0.029 &  -0.180$\pm$0.018 &   -4.0$\pm$1.1 &  207.3$\pm$2.2 &  -13.0$\pm$1.9 & 1122 \\
 NGC 1342 &  -8.581$\pm$0.058 &   0.271$\pm$0.027 &  -0.176$\pm$0.018 &   20.5$\pm$1.3 &  217.8$\pm$2.5 &    1.0$\pm$2.8 &  452 \\
 NGC 1647 &  -8.517$\pm$0.052 &  -0.004$\pm$0.000 &  -0.156$\pm$0.016 &   18.6$\pm$0.4 &  224.0$\pm$0.8 &    2.7$\pm$1.0 &  144 \\

\hline
\end{tabular} }
\end{table*}

From the entries in Table 15, we note here that the $\alpha$-enhanced metal-rich clusters (thin disc) have Galactic rotation velocities (V$_{GSR}$) intermediate to those of metal-poor high-$\alpha$ clusters (thick disc) and young metal-rich low-$\alpha$ clusters (thin disc; for example, IC 2391,  IC 2602) and are  more similar to the $\alpha$-enhanced ([$\alpha$/Fe]=\,0.0 to 0.2 dex), metal-rich ([Fe/H]=\,-0.2 to 0.2 dex) older field stars detected in the solar vicinity (see Figures 1 \& 4 in Adibekyan et al. 2013; Bensby et al. 2014). Although the origin of such objects is still in debate, Adibekyan et al. (2011, 2013) have suggested a possible origin of these $\alpha$-enhanced metal-rich stars in the inner Galaxy (bulge) followed by a subsequent migration to the solar vicinity. 
 
Bulge and thick disc dwarfs (Bensby et al. 2010) and giants (Alves-Brito et al. 2010) are indistinguishable in [$\alpha$/Fe] versus [Fe/H] trends, with a star-to-star scatter of only 0.03 dex over the range of $-1.5$ to $+0.3$ dex in metallicity. And the bulge sample with [Fe/H]$<\,-$0.5 dex is dominanted by stars older than 10-12 Gyr while for [Fe/H]$>\,-$0.5, the bulge sample is represented by stars covering a wide range in age from a few Myr to 13 Gyr (Bensby et al. 2013). 

The most $\alpha$-enriched ([$\alpha$/Fe]$>$\,0.15$\pm$0.04) thin disc OCs in the present sample have R$_{\rm gc}$ and age, respectively, as follows: 8.2 kpc, 8.3 Gyr (NGC 6791), 8.9 kpc, 6.9 Gyr (NGC 7142), and 9.0 kpc, 7.1 Gyr (NGC 188). We suggest by the striking chemical similarity between the old thin disc OCs (Figure \ref{alpha_ageoc}) and bulge giants that these clusters might have formed close to the bulge or in the inner disc and migrated radially out to the solar vicinity or the orbital eccentricity of these clusters could have helped in reaching them to present locations in the Galactic disc. 
In this vein, we note that Jilkova et al. (2012) trace orbits of a metal-rich OC NGC 6791 for an age equal to its age which resulted in a low probability that NGC 6791 was formed close to the bulge (R$_{\rm gc}$\,=\,3$-$5 kpc). But, as noted by Carraro (2014), the orbital eccentricity (e$=$0.3) of the cluster aids in reaching it present locations in the immediate solar neighbourhood from the birthplace in the inner disc.

\section{Radial abundance distribution of OCs and the field stars}
The radial abundance gradient (and azimuthal variations) offers a challenge to models of galaxy formation and evolution. The gradients defined by OCs have been the subject of previous studies. The current picture suggests a gradient for [Fe/H] of about $-$0.06 dex kpc$^{-1}$ (Friel et al. 2002; Pancino et al. 2010) to $-$0.20 dex kpc$^{-1}$ (Frinchaboy et al. 2013) in the radial range 5 to 10 kpc with a nearly flat trend ($-$0.02 dex kpc$^{-1}$) over the entire radial extent of the disc at R$_{\rm gc}>$12 kpc, with a break occurring around 11 kpc (Sestito et al. 2008; Pancino et al. 2010; Yong et al. 2012; Frinchaboy et al. 2013).  

We show in figure \ref{gradient}, the run of [Fe/H] versus R$_{\rm gc}$ (top panel) and [$\alpha$/Fe] versus R$_{\rm gc}$ (bottom panel) for the present homogeneous sample of 79 OCs (a total of 28 OCs from our study and 51 OCs from the literature). We have, by recalculating the R$_{\rm gc}$ of each cluster using R$_{\odot}$ of 8.0$\pm$0.6 kpc (Ghez et al. 2008), placed clusters on a common distance scale. Thus, our final sample of OCs cover a range of 5.0 to 24.0 kpc in R$_{\rm gc}$, $-$0.5 to 0.3 dex in metallicity and an age of few Myr to 9 Gyr. To our knowledge, we are the first to provide a  homogeneous high-resolution abundance analysis of elements from Na to Eu for twenty-eight OCs in the radial range R$_{\rm gc}$=\,7.7 to 11.3 kpc which constitute about 35\% of OCs explored so far with high-resolution spectroscopy. Before we use this homogeneous sample to investigate the radial abundance distribution in the disc, it is worth while to know the vertical and the age distribution of OCs with increasing R$_{\rm gc}$ in the Galactic disc.

\begin{figure}
\begin{center}
\includegraphics[trim=0.2cm 6.3cm 7.8cm 4.2cm, clip=true,height=0.27\textheight,width=0.50\textwidth]{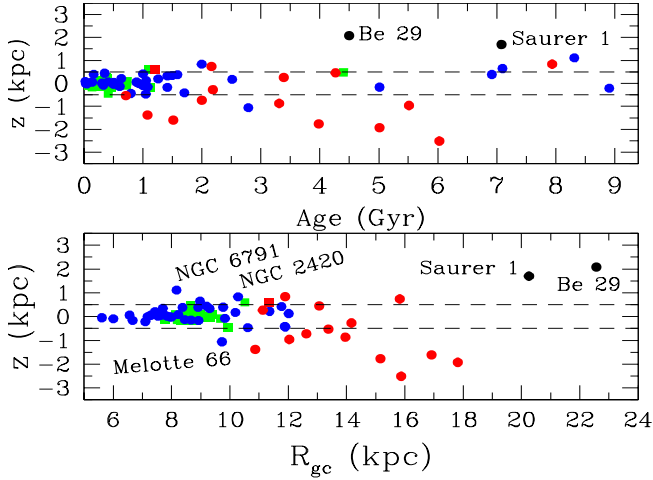}
\caption[]{The distance of the cluster away from the Galactic plane as a function of cluster's age (top panel) and R$_{\rm gc}$ (lower panel). All the symbols have their usual meaning as in Figure \ref{gradient}. The horizontal dashed lines mark the bounds of $\pm$0.5 kpc in z. }
\label{rgc_age}
\end{center}
\end{figure}

We show in Figure \ref{rgc_age}, the distance of the cluster away from the Galactic midplane as a function of cluster's age (top panel) and R$_{\rm gc}$ (lower panel). The great majority of clusters with ages younger than that of Hyades (with age 0.8 Gyr), as noted previously in studies by Phelps et al. (1994) and Chen et al. (2003), are uniformly distributed close to the galactic plane in the immediate solar vicinity. Close inspection of figure \ref{fe_ageoc} (top panel) and figure \ref{rgc_age}, indicate that the cluster sample populating the radial extent beyond 11 kpc are older with ages from 1.0 to 8 Gyr and metal-poor with kinematics typical of the thick disc and halo populations. The region inward of 11 kpc is populated completely with thin disc clusters younger than 1.5 Gyr while five thin disc clusters, namely NGC 6253 (6.7 kpc, 5.0 Gyr), Collinder 261 (7.2 kpc, 9.0 Gyr), NGC 6791 (8.2 kpc, 4.4 Gyr), NGC 7142 (8.9 kpc, 6.9 Gyr), and NGC 188 (9.0 kpc, 7.1 Gyr) are older and none belong to the thick disc. The lack of older OCs in the solar vicinity and farther away from the Galactic plane (z$\,>\,\pm$0.5 kpc) has been related to the severe destruction of these clusters with frequent encounters with the giant molecular clouds over their lifetime (Lamers \& Gieles 2006). 

All but three OCs with kinematics typical of the thick disc and halo populations in the radial range 11 to 23 kpc are metal-poor by [Fe/H] $<-$0.2 dex (Figure \ref{gradient}) and are located at relatively large heights above the Galactic plane (0.5$\,<\lvert\,z\rvert<\,$2.5 kpc) while almost all the thin disc clusters fall within the bounds of $\lvert\,z\rvert<\,$0.5 kpc (Figure \ref{rgc_age}). Therefore, attempts to measure the radial gradients using the OCs distributed in the outer disc (R$_{\rm\,gc}>\,$11 kpc) will be severely influenced by the vertical gradients whose affect is minimal for the sample in the vertical slice $\lvert\,z\rvert<\,$0.5 kpc and thus unimportant for the radial gradient inferred for OCs in the radial range 6 to 11 kpc (Cheng et al. 2012; Hayden et al. 2014). 

We measure a radial metallicity gradient of $-$0.052$\pm$0.011 dex kpc$^{-1}$ using a linear fit to the OC sample falling in the radial range 6 to 12 kpc and for $\lvert\,z\rvert<\,$0.5 kpc which is consistent with such gradients reported previously for OCs in Friel et al. (2002; $-$0.06$\pm$0.01 dex kpc$^{-1}$) and Pancino et al. (2010; $-$0.06$\pm$0.01 dex kpc$^{-1}$), and for the Galactic Cepheids (Luck \& Lambert 2011; $-$0.062$\pm$0.002 dex kpc$^{-1}$). Unlike the inner disc sample, the number of clusters with measured metallicities at R$_{\rm\,gc}>\,$12 kpc is very small with a large spread in age and all of them occupy the vertical bin 0.5$\,<\lvert\,z\rvert<\,$2.5 kpc, the derived gradient of $-$0.015$\pm$0.007 dex kpc$^{-1}$ is heavily weighted by a few clusters and is not truly a representative of metallicity variation with increasing radius in the midplane of Galactic disc. Previous studies have also shown evidence that the radial gradient in the inner disc turns to a flat distribution at large Galactocentric radii with transition occurring around 12 kpc (Magrini et al. 2009; Friel et al. 2010; Yong et al. 2012; Frinchaboy et al. 2013). 

\begin{figure}
\begin{center}
\includegraphics[trim=0.1cm 8.2cm 7.7cm 4.5cm, clip=true,height=0.19\textheight,width=0.48\textwidth]{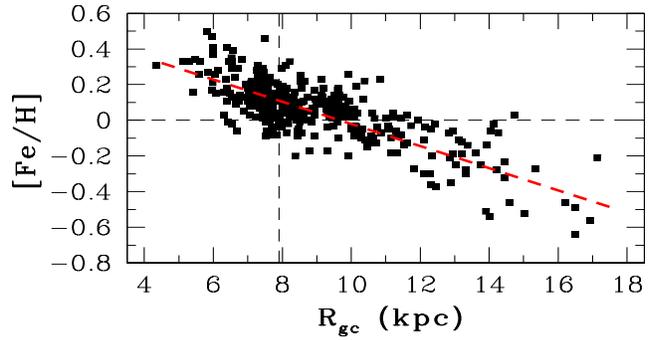}
\caption[]{[Fe/H] vs. R$_{\rm gc}$ for a sample of Cepheids analysed by Luck \& Lambert (2011) where the R$_{\rm gc}$ of a Cepheid is calculated using (R$_{\odot}$=\,7.9 kpc). A fit to the data with a line of slope $-$0.062 dex kpc$^{-1}$ is represented as a broken red dashed line. }
\label{gradient_cepheids}
\end{center}
\end{figure}

However, the shallow radial gradient seen for OCs at large radii is not observed in the analyses of a large sample of F and G field dwarfs from the SEGUE survey (Cheng et al. 2012), field giants in the APOGEE survey (Hayden et al. 2014) and for Cepheids (see, for example, Figure \ref{gradient_cepheids}) over the whole radial range of 4 to 18 kpc but for stars sampled in the vertical bin of $\lvert\,z\rvert<\,$0.5 kpc. Both Cheng et al. (2012) and Hayden et al. (2014) have measured a steep radial gradients of about $-$0.06 dex kpc$^{-1}$ (0.15$\,<\lvert\,z\rvert<\,$0.5 kpc) and $-$0.087$\pm$0.002 dex kpc$^{-1}$ (0.0$\,<\lvert\,z\rvert<\,$0.25 kpc), respectively, close to the plane and have confirmed that the radial gradients become flatter as one moves away from the Galactic midplane in the vertical range 0.5$\,<\lvert\,z\rvert<\,$2 kpc. Again the sample of Cepheids used in measuring the radial abundance gradients is concentrated close to the Galactic plane within $\lvert\,z\rvert<\,$0.5 kpc (Luck \& Lambert 2011). 
At R$_{\rm\,gc}>\,$12 kpc and away from the midplane, the mean metallicity of about $-$0.3 dex and the flat radial gradient observed for the present sample of OCs is consistent with very similar values reported for field dwarfs (Cheng et al. 2012) and field giants (Hayden et al. 2014). These results suggest that the OCs and field stars yield consistent radial gradients if the comparison samples are drawn from the similar vertical slices. 
Claims of a difference in radial gradients between clusters and field stars at R$_{\rm\,gc}>\,$12 kpc (Cheng et al. 2012) result from comparison of samples drawn occupying different scale heights ($\lvert\,z\rvert<\,$0.5 kpc for young OCs and field stars versus $\lvert\,z\rvert>\,$0.5 kpc for the older outer disc OCs). 

Noting the change in radial gradient measured for field stars lying in each $\lvert\,z\,\rvert$ slice (Cheng et al. 2012; Hayden et al. 2014), the abrupt break in the declining metallicity of OCs with increasing radius at R$_{\rm\,gc}$ of 12 kpc and a further flattening out to the entire radial extent could be the result of the superposition of the measured negative gradient for young clusters at small z with a flat gradient derived for older OCs spanning a wide range in heights above the Galactic midplane. The present sample of OCs in the radial range 6 to 12 kpc trace the radial metallicity gradient close to the disc midplane while the older OCs at R$_{\rm\,gc}>\,$12 kpc represent the metallicity gradient in the disc high above the disc midplane.

In figure \ref{gradient}, we present the variation of [$\alpha$/Fe]-ratios with increasing R$_{\rm gc}$ and have measured radial gradients of $+$0.014$\pm$0.005 dex kpc$^{-1}$ (R$_{\rm gc}=$6$-$12 kpc) and 0.00$\pm$0.005 dex kpc$^{-1}$ (R$_{\rm gc}>\,$ 12 kpc). There is an indication that the outer disc OCs found to have slightly higher mean [$\alpha$/Fe] values relative to inner disc clusters (Carraro et al. 2004; Yong et al. 2012) which as demonstrated clearly in the figure \ref{gradient} that the $\alpha$-enrichment is principally governed by the population of thick disc OCs in the outer Galactic disc. 

\begin{figure}
\begin{center}
\includegraphics[trim=0.2cm 5.0cm 2.4cm 4.2cm, clip=true,height=0.34\textheight,width=0.68\textwidth]{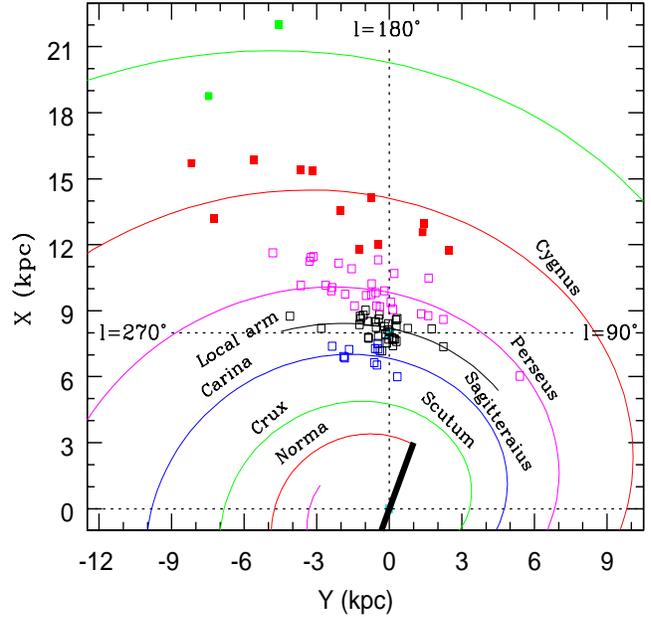}
\caption[]{The spatial distribution of OCs projected onto the Galactic plane along with the log-periodic spiral arm pattern with an inward pitch angle of $p$ = 13$^\circ$ (Bobylev \& Bajkova 2014) and the local arm pattern with $p$ = 12.8$^\circ$ is drawn from Reid et al. (2014). The Galactic center and the Sun are located at (0,0) kpc and (8,0) kpc, respectively. OCs occupying different spiral arms are represented in color same as that of the arm. The Galactic bar, inclined at an angle of 20$^\circ$ to the line joing the Sun to the Galactic centre, is shown as a thick black line.}
\label{spiral_oc_position}
\end{center}
\end{figure}

The study of Luck \& Lambert (2011) suggests a lack of azimuthal variation in [Fe/H] of Cepheids drawn from several azimuthal cuts having a depth of 1 kpc each and lying at different Galactocentric radius. The sample of clusters in this paper does not allow us to draw conclusions on the azimuthal variation in [Fe/H] of OCs; the total sample of 79 OCs shows that about 58 clusters are located closer to the Galactic plane. Of these 58 OCs, thrice as many OCs are located in the third galactic quadrant than in the second and the data is concentrated more in the radial than in azimuthal direction (Figure \ref{spiral_oc_position}). In a recent study using the photometric and spectroscopic metallicities for 172 clusters compiled from various literature papers and metallicities of Cepheids drawn from Genovali et al. (2014), Netopil et al. (2016) trace a little evidence for azimuthal dependence of metallicity, in concert with the results of previous studies (Luck \& Lambert 2011; Genovali et al. 2014). 
All but the new sample of clusters analysed in this paper are included in the analysis of Netopil et al. (2016). The addition of these twelve clusters to Netopil et al. (2016) sample add no scatter but endorse their conclusions.

The current homogeneous sample of 79 OCs allow us to conclude that the variation in [Fe/H] with increasing Galactocentric radius is steeper close to the Galactic midplane; at R$_{\rm\,gc}<\,$12 kpc, all the sampled clusters are younger than 1.5 Gyr with thin disc kinematics, close to the plane ($\lvert\,z\rvert<\,$0.5 kpc) and represent a steep slope in the radial range 6 to 12 kpc. At R$_{\rm\,gc}>\,$12 kpc, almost all the sampled clusters 
are from the thick disc with a striking spread in age and height above the midplane (0.5$\,<\lvert\,z\rvert<\,$2.5 kpc) and represent a shallow gradient over the entire radial extent of the outer Galactic disc. Stitching together these two regions results a constant decline in [Fe/H] with increasing R$_{\rm\,gc}$ with a sudden change in slope at 12 kpc and a further flattening out to the entire radial extent of the Galactic disc as observed today. Such behaviour of radial metallicity distribution of OCs in the Galactic disc is endorsed by the field stars and Cepheids, where the field stars and Cepheids for $\lvert\,z\rvert<\,$0.5 kpc trace a constant steep decline of metallicity while away from the Galactic midplane (0.5$\,<\lvert\,z\rvert<\,$2 kpc), field stars represent a shallow gradient over the entire radial extent of the disc. Therefore, we argue that the OCs, field stars and Cepheids yield consistent radial gradients if the comparison is limited to samples drawn from the similar vertical heights.

\subsection{Comparison with chemodynamical models}
The chemodynamical simulations of Minchev et al. (2013) predict a steep gradient throughout the radial range 5 to 16 kpc for samples of stars younger than 2 Gyr drawn from any slice in the range 0$\,<\lvert\,z\rvert<\,$3 kpc. They also found a signature of flattening of the radial gradient due to radial migration induced by the strong interation of stars older than 2 Gyr with the bar, but the young population is hardly affected out to R$_{\rm gc}<$ 12 kpc, where some flattening occurs (see figure 5 in Minchev et al. 2013). As a result, the initial and the present-day gradient derived for the younger populations are very similar out to 12 kpc. This fairly explains the similarity in the radial gradients observed for the younger tracers such as OCs of age less than 2 Gyr (Figure \ref{gradient} and figure \ref{rgc_age}), field stars (Cheng et al. 2012; Hayden et al. 2014) and Cepheids (Luck \& Lambert 2011; Genovali et al. 2014) for R$_{\rm gc}<$ 12 kpc. The steep gradient of slope $-$0.058 dex kpc$^{-1}$ for stars younger than 2 Gyr in the radial range 6$-$11 kpc and a shallow gradient of old stellar populations predicted after taking the radial migration into account is in fair agreement with our values of gradient measured for respective age groups.

In Minchev et al.'s simulation, the [O/Fe] profiles (proxy for [$\alpha$/Fe]) have been associated with a very weak flattening over time and a slight enrichment with increasing radii and the same is reflected in the $\alpha$-enrichment observed for OCs (bottom panel of figure \ref{gradient}). These results further strengthen, as noted previously by Phelps et al. (1994) and Chen et al. (2003), that the sample of outer disc clusters has been subjected to significant selection effects; they are readily observed if they are away from the plane due to high interstellar extinction in optical wavelengths at low latitudes in the direction of Galactic anti-centre. Again they are likely to be the older, as the young OCs have yet to move out from their birthplaces in Galactic midplane and, thus, obscured by heavy extinction caused by about 90\% of the absorbing material lying within -5$<$\,b\,$<$5 deg of the Galactic plane (Joshi 2005) which raises the level of difficulty in observing them in visible bands.

Before turning our attention to the uniform chemical content displayed by the outer disc OCs over the entire radial range at R$_{\rm gc}>\,$12 kpc, we revisit the metallicity distribution functions (MDFs) of the field red giants from APOGEE survey. Hayden et al. (2015) showed that field giants in the Galactic plane ($\lvert\,z\rvert<\,$0.5 kpc) have a significant radial gradients over the entire radial extent of the disc: the peak of the Milky Way’s MDF shifts from being metal-rich ([Fe/H]=+0.32 for 3$<\,R_{\rm gc}\,<$5 kpc) in the inner Galaxy through roughly solar in the solar vicinity ([Fe/H]=+0.02 for 7$<\,R_{\rm gc}\,<$9 kpc) to metal-poor [Fe/H]=$-$0.48 (13$<\,R_{\rm gc}\,<$15 kpc) in the outer disc (see Figure 5 in Hayden et al.). But at $\lvert\,z\rvert\,>$0.5 kpc, the MDFs constructed from samples of stars drawn from different radial bins have peaked around [Fe/H]=$-$0.3 dex, suggesting that the radial metallicity gradient is flat over the entire radial extent of the Galactic disc away from the midplane. 

The MDFs observed in Hayden et al. (2015) are realized in the N-body+smooth particle hydrodynamics simulations of Loebman et al. (2016) where they have shown that migrated stars represent a large fraction ($>\,$50\%) of stars in the outer disc. They show clearly in their simulation that stars formed on a short timescale (8.8$\pm$0.6 Gyr) in a relatively small region of well-mixed gas in the inner Galaxy could migrate out to their present locations, all over the Galaxy leading to apparently chemically homogeneous (very similar iron abundance) sample of stars over the entire radial extent of the disc at $\lvert\,z\rvert\,>$0.5 kpc. None of the outer disc clusters in this study are distinguishable with their iron abundances. We suggest from the striking chemical similarity between outer disc OCs and field stars from APOGEE (Hayden et al. 2015) that these OCs would have undergone radial mixing (Sellwood \& Binney 2002; Ro\v{s}kar et al. 2008; Sch\"{o}nrich \& Binney 2009; Loebman et al. 2016) and subsequently moved to current locations from their birth-sites in the inner disc enriched with very similar iron content.

\section{Computation of OC birth-sites}
To explore the feasibility of radial mixing and the possibility that the outer disc OCs have formed in the inner disc from a well-mixed reservoir of gas and have moved away from their birthplaces to their present locations, we have studied the dynamics of OCs under the influence of a multicomponent Galactic gravitational potential. We have adopted the static, axisymmetric Galactic gravitational potential model of Flynn, Sommer-Larsen \& Christensen (1996, hereafter FSC96) representing a spherical central bulge composed of a superposition of two Plummer spheres (Plummer 1911), a disc potential made up of three concentric analytical discs of Miyamoto-Nagai (1975), and a massive dark halo modeled as a spherical logarithmic potential. For this potential, we adopt a distance of the Sun to Galactic centre of 8 kpc with a local circular velocity of 220  km
s$^{-1}$ (same as adopted earlier in Wu et al. 2009). 

The orbit of a cluster under the influence of a galactic potential is commonly described as a superposition of the main circular motion that defines the guiding radius and harmonic oscillations (epicycles) about the circular orbit. Following Sch\"{o}nrich \& Binney (2009), radial mixing occurs through blurring, where a star's epicycle amplitude increases with time by scattering at an orbital resonance or by non-resonant scattering by molecular clouds without changing its angular momentum, and churning, where a star upon gaining (losing) the angular momentum in a scattering event that preserves the overall distribution of angular momentum moves outwards (inwards) while maintaining a circular orbit, i.e., churning does not  add random motion and heat the disc radially. In contrast, blurring conserves the angular momentum of individual stars but heats the disc in the radial direction, thereby configuring the orbit to eccentric over time so that a star can cover a wider range in galactocentric radii. As a result, all the previously used static and axisymmetric Galactic potential models of Fellhauer et al. (2006), Allen \& Santillan (1991) and FSC96, and Bovy et al. (2015) in the orbit reconstruction of OCs by Vande Putte et al. (2010), Wu et al. (2009), and Cantat-Gaudin et al. (2016), respectively, and the one adopted in this study yield spurious birthplaces if the radial mixing happened by churning. Therefore, this process is successful in recovering the genuine orbital history and birthplace of OCs only if they underwent radial mixing originating from blurring process. 

The relevant code for the integration of cluster orbits has been kindly provided by Zhen-Yu Wu (private communication). It was used previously in their analysis of kinematics and orbits for a sample of 488 OCs (Wu et al. 2009). The equations of motion constructed in cylindrical coordinates R, $\phi$ and z, where R is the galactocentric radius, $\phi$ is the azimuthal angle and z is the distance of the cluster away from the Galactic plane, with the adopted potential conserves both the energy and angular momentum at a level of 10$^{-14}$ and 10$^{-12}$, respectively, over the whole orbit integration. 

Hamilton's equations, representing a set of first order differential equations were integrated using the Bulirsch-Stoer algorithm of Press et al.(1992) implemented in FORTRAN code with adaptive time-steps. Starting with a cluster's current position and velocity components referenced to the Galactic standard of rest (Table 15), we followed the OC's trajectory backward in time over a period of 5 Gyr to ensure that even the youngest clusters with an age of few Myr could complete sufficient galactic orbits so that the averaged orbital parameters can be determined with fair certainty. But for each cluster, the birthplace and z-component at birth were computed by integrating the orbit backward in time for an interval equal to its age. 

The dispersions for computed orbital parameters have been evaluated following Dinescu et al. (1999; see also Wu et al. 2009), and generated a set of 1000 initial conditions for each cluster by adding Gaussian random errors to the observed absolute proper motions, RVs and distances from the Sun which are used in calculating positions and space velocity components of the cluster. The standard deviations are taken to be the errors associated with the input proper motions and RVs, while we add errors of 10\% each in distance and ages. The final error associated with each of the averaged orbital parameters have been calculated by running 1000 separate integrations for each cluster.

\begin{table*} 
\flushleft { {\bf Table 16.} The relevant orbital parameters and their errors of 79 OCs in our sample calculated with the Galactic gravitational model of Flynn, Sommer-Larsen \& Christensen (1996). All the columns are self-explanatory (see the text for reference). The full table is available with the online version of the paper.}

{\fontsize{6}{7}\selectfont
\begin{tabular}{lcccccccccc} \hline
\multicolumn{1}{l}{Cluster}& \multicolumn{1}{c}{R$_{\rm a}$}& \multicolumn{1}{c}{R$_{\rm p}$}& \multicolumn{1}{c}{R$_{\rm gc}$} & \multicolumn{1}{c}{R$_{\rm birth}$}& \multicolumn{1}{c}{z$_{\rm birth}$} & \multicolumn{1}{c}{z$_{\rm max}$} & \multicolumn{1}{c}{eccentricity} & \multicolumn{1}{c}{T$_{\rm p}$} & \multicolumn{1}{c}{T$_{\rm z}$} & \multicolumn{1}{c}{J$_{\rm z}$}  \\
\multicolumn{1}{l}{ }& \multicolumn{1}{c}{(kpc)}& \multicolumn{1}{c}{(kpc)}& \multicolumn{1}{c}{(kpc)}& \multicolumn{1}{c}{(kpc)}& \multicolumn{1}{c}{(kpc)} & \multicolumn{1}{c}{(kpc) } & \multicolumn{1}{c}{ } & \multicolumn{1}{c}{(Myr)} & \multicolumn{1}{c}{(Myr)} & \multicolumn{1}{c}{(kpc km s$^{-1}$)}  \\
\hline

\multicolumn{11}{c}{\bf  \normalsize {Thin disc} } \\
   
 NGC 752 &8.39$\pm$0.03 &7.25$\pm$0.08 &8.31$\pm$0.05 &8.33$\pm$0.06 &0.15$\pm$0.09 &0.26$\pm$0.02 &0.07$\pm$0.01 & 217.4$\pm$0.4 &45.8$\pm$0.3 &-1721.0$\pm$8.2 \\
 NGC 1342 &8.79$\pm$0.01 &8.00$\pm$0.04 &8.59$\pm$0.07 &8.51$\pm$0.41 &0.03$\pm$0.08 &0.17$\pm$0.01 &0.05$\pm$0.00 & 230.5$\pm$0.6 &47.1$\pm$0.6 &-1874.2$\pm$6.4 \\
 NGC 1647 &9.00$\pm$0.03 &8.06$\pm$0.03 &8.52$\pm$0.05 &8.32$\pm$0.63 &0.16$\pm$0.02 &0.16$\pm$0.01 &0.06$\pm$0.00 & 233.9$\pm$0.7 &47.6$\pm$0.4 &-1908.1$\pm$7.6 \\

\hline
\end{tabular} }
\end{table*}

Entries in Table 16 list the averaged orbital parameters: the apogalacticon (R$_{a}$), perigalacticon (R$_{p}$), R$_{\rm gc}$, birthplace (R$_{birth}$), distance of the cluster away from the Galactic plane at birth (z$_{birth}$), the maximum vertical amplitude (z$_{max}$),  orbital eccentricity defined by e$=$(R$_{a}$-R$_{p}$)/(R$_{a}$+R$_{p}$), orbital period (T$_{p}$), epicyclic period (T$_{z}$) and J$_{z}$. The negative (positive) values of J$_{z}$ represent the prograde (retrograde) motion of the cluster in (against) the sense of Galactic rotation.

\begin{figure}
\begin{center}
\includegraphics[trim=0.1cm 4.2cm 2.5cm 4.5cm, clip=true,width=0.68\textwidth,height=0.36\textheight]{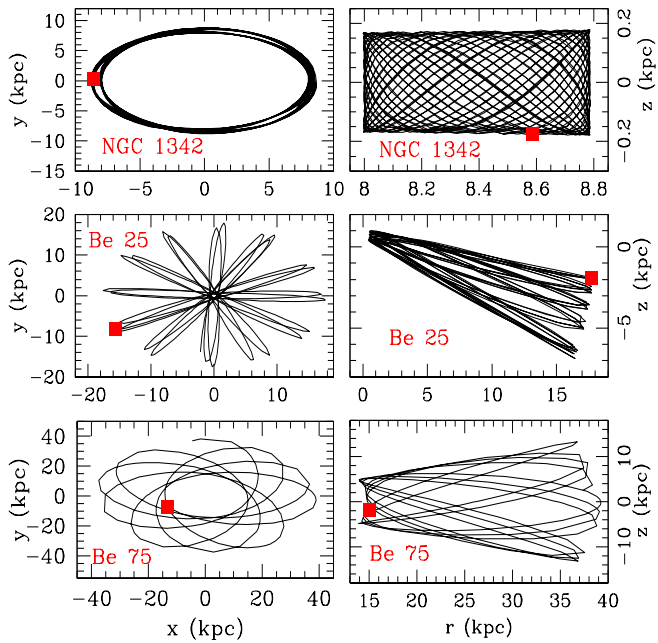}
\caption[]{Orbits of OCs projected onto the Galactic plane (left panel) and in the meridional plane (right panel) computed for a time interval of 5 Gyr for NGC 1342, Berkeley 25 and Berkeley 75. The filled square (red) represent the current observed position for each cluster.}
\label{oc_orbit} 
\end{center}
\end{figure}

We notice that all the OCs with the exception of Be 25 follow symmetric quasi-periodic orbits with respect to the Galactic plane, an example of which is shown in Figure \ref{oc_orbit} for NGC 1342. Be 25 (5 Gyr) follows an irregular and asymmetric orbit about the Galactic plane, and may represent a perturbed orbit, for which it is difficult to recover the orbital history and a genuine birthplace. Another possibility would be that the input data (mostly distances and proper motions) used in the orbit computations of Be 25 are quite erroneous (see, for example, Wu et al. 2009; Cantat-Gaudin et al. 2016 for a discussion of the impact of proper motions on orbit computations).

It is evident from the entries in Table 16 that most of the young clusters (ages$\,<$ 2 Gyr) have circular orbits lying close to the Galactic midplane (i.e., a low z$_{max}$ during the integration time of 5 Gyr) where they were born with their velocity aligned with the local circular velocity. Whereas clusters with kinematics typical of the older Galactic populations have relatively eccentric orbits and therefore 
span a wide range in R$_{\rm gc}$ and also make large excursions away from the Galactic midplane (i.e., a larger $z_{max}$). As a result the thick disc clusters spend most of their lifetimes far away from the Galactic midplane, a pleasing result in support of their survival for longer periods of time. However, the survival of a few old clusters with thin disc kinematics, namely NGC 2682 (4.3 Gyr), NGC 6791 (4.4 Gyr), NGC 6253 (5 Gyr), NGC 7142 (6.9 Gyr), NGC 188 (7.1 Gyr) and Collinder 261 (9.0 Gyr), while keeping their orbits close to the Galactic plane suggest that they might have been the most massive OCs when they were formed. 

Our investigation reveals that almost all the thin disc OCs have birthplaces very close their current locations (similar R$_{\rm gc}$ but a different azimuth; see columns 4 and 5 in Table 16). In contrast, the estimated birthplaces of all but three outer disc OCs fall close to the Galactic plane in the radial range 9 to 12 kpc but with a relatively large dispersion in R$_{birth}$ and z$_{birth}$ caused by the observational uncertainties in proper motions and distances from the Sun. While one thick disc cluster Be 75 (Figure \ref{oc_orbit}) and the two clusters, namely Be 29 and Saurer 1, with halo membership have formation sites around 30 kpc and the respective perigalacticon in three not ever reach closer than 14 kpc to the Galactic center. The low surface gas density of the Milky Way Galaxy at large radii for efficient star formation (Yin et al. 2009) may put in doubt their origin in the outer Galactic disc. A likely suggestion, as previously made by Cantat-Gaudin et al. (2016), would be that these clusters were perturbed soon after their formation in the inner disc or have their measured proper motions quite erroneous owing to great distances and lack of cluster members brighter than V=15 mag (Cantat-Gaudin et al. 2016). 

In this regard, the ongoing GAIA mission, aims at providing accurate distances, proper motions and RVs of about one billion stars down to V=20 (Perryman et al. 2001) will, therefore, be useful to confirm or deny the association of these clusters to the Galactic disc stellar populations. The predicted post-launch sky-averaged end-of-mission proper motion errors\footnote{\url{http://www.cosmos.esa.int/web/gaia/science-performance}} of M stars range from 5$\,\mu$as at V=15 to 70$\,\mu$as at V=20 with quite significant RV erorrs of 15 km s$^{-1}$ in the V=15-16 range for red giants. The insufficient RV precision of GAIA will prevent accurate kinematic studies of OCs but will make it easier to identify the most probable cluster members for follow-up high-resolution spectroscopy. 

A possibility exists that OCs in the outer Galaxy have been accreted. We note that Law \& Majewski (2010) acknowledged the similarity in age, metallicity, luminosity, and structural properties of Be 29 and Saurer 1 with some globular clusters in dwarf spheroidal galaxies (dSph) and suggested that these two OCs may belong to a family of sparse globular clusters formed in dSphs, stripped from their birth galaxies, and subjected to the gravitational tides of the Milky Way. Again the positive latitudes of Be 29 and Saurer 1 in the Northern Hemisphere when all other distant OCs have negative latitudes (Figure \ref{rgc_age}), consistent with the southward bend of the outer disc (Levine et al. 2006), and the Halo membership probabilities may put in doubt their origin in the Milky Way. At the present time, the quality of the proper motions and distances from the Sun does not allow us to draw firm conclusions on the origin of Be 29 and Saurer 1.

Although our results do not support L\'{e}pine et al. (2011)'s extreme interpretation of two zones in the radial distribution of [Fe/H] with flat slopes separated by a step-like discontinuity of height $\Delta$[Fe/H]=\,0.3 dex at 8.5 kpc (see Figure 4 in Lepine et al. 2011), but endorse their suggestion that the OCs at R$_{\rm gc}$ of 15 to 20 kpc were born inside a radius of 12 kpc and have travelled to their present positions. It has further been described pictorially in L\'{e}pine et al. that the effective galactic potential curves (Figure 2 in their Paper) become shallower at large radii so that a transient energy perturbation about the circular orbit provides the cluster with an extreme coverage in Galactic radii. As the radial mixing operates at a slow pace, its affect is more visible in the older Galactic populations than in the younger ones. Therefore, the oldest outer disc OCs have enough time to have been influenced by scattering at orbital resonances caused by the spiral density waves. 

The proximity of the resonances of spiral pattern with the disc material at about 11 kpc provides a further support to our assertions. For any given rotation curve, the location of the inner and outer Lindblad resonances (ILR, OLR) and corotation tightly depends on the the pattern speed ($\omega_{p}$) of spiral arms for which values ranging from 10 to 30 km s$^{-1}$ kpc$^{-1}$ are available in the literature (Naoz \& Shaviv 2007). Naoz \& Shaviv (2007)'s study of the Milky Way spiral arm kinematics using the birthplace of OCs find evidence for multiple spiral sets, supporting the superposition of m\,=\,2 and m\,=\,4 armed patterns of the Milky Way (L\'{e}pine et al. 2001; Englmaier et al. 2008). The faster Carina arm and the Orion arm, nealy corotating with the solar system, have $\omega_{p}$ of about 29 km s$^{-1}$ kpc$^{-1}$ that places the the resonances at: R$_{\rm CR_{2}}$\,=\,7.9 kpc, R$_{\rm OLR_{2}}$\,=\,13.6 kpc for m=2 (2:1 resonances) and R$_{\rm ILR_{4}}$\,=\,4.8 kpc and R$_{\rm OLR_{4}}$\,=\,10.9 kpc for m=4. Whereas the slower Perseus arm with $\omega_{p}$ of about 20 km s$^{-1}$ kpc$^{-1}$ situates the resonances at: R$_{\rm CR_{4}}$\,=\,11.5 kpc, R$_{\rm ILR_{4}}$\,=\,8.0 kpc and R$_{OLR_{4}}$\,=\,15.6 kpc for m=4 and R$_{OLR_{2}}$\,=\,20.0 kpc for m=2. 

It has been shown in many N-body simulations of isolated spiral discs that spiral arms are transient, recurring, co-rotating features
with the stellar disc stars at all radii that, because of their ubiquitous corotation resonances cause more radial migration than density
waves (Sellwood \& Binney 2002; Sellwood 2011; Grand \& Kawata 2015; Fujii \& Baba 2012). The larger changes in angular momentum of stars (churning) always occur near the corotation resonance that leads to the radial migration of stars of few kpc from their birth radii on a short period, while the effect at the inner/outer Lindblad resonances (blurring) is smaller (Sellwood \& Binney 2002; Minchev \& Famaey 2010). Through N-body simulations, Fujii \& Baba (2012) find that clusters lose (gain) at most 50\% of their angular momentum within 1 Gyr. They found a timescale of 100 Myr for OC migration and suggested that OCs in the Galactic disc older than 100 Myr could have migrated radially about 1.5 kpc from their initial birth radii. If such a violent mechanism is operating in the Galactic disc, it could wipe away the radial gradient traced from the present sample of younger OCs ($<$ 2 Gyr) in a few 100 Myr. Hence, an alternative explanation is needed for the preservation of observed negative gradient in the inner disc. 

Recently Grand \& Kawata (2015) proposed a mechanism that is less pronounced and qualitatively different from the significant radial migration localized to corotation radius of the density wave-like spiral arm. They show, using N-body simulations of a giant disc galaxy, that the spiral arms are transient, recurring features (100 Myr) whose pattern speeds decrease with radius, in such a way that the pattern speed is almost equal to the rotation curve of the galaxy. As a result, stars continually feel the tangential force of the spiral arm until it disrupts at about 100 Myr; stars on the trailing (leading) side are torqued radially outward (inward) and gain (lose) angular momentum. As the spiral stellar density enhancement corotates with stars at every radius, the continual angular momentum gains/loses of these migrating stars leads to streaming motions along the both sides of the spiral arm. They show clearly in their simulations that the radial migration acts to maintain the negative radial metallicity gradient but significantly broadens the metallicity distribution at every radius by about 0.8 dex (see Figure 2 in Grand \& Kawata 2015). The motion of star clusters in spiral arms and the angular momentum exchange is similar to that of stars investigated in simulations of stellar discs (Fujii \& Baba 2012; Grand et al. 2012). 

We suggest from the lack of significant flattening and large dispersion at every radii in the radial metallicity distribution of OCs that the radial migration by churning may not be as efficient as the model predictions from Fujii \& Baba (2012) and Grand \& Kawata (2015). Recent N-body smooth particle hydrodynamics simulations of Kubryk et al. (2015) track the impact of stellar migration (blurring+churning) on the chemical evolution of the Milky Way disc. Their model predicts that observed gradients will be flatter for older populations, as migration could wipe away the gradients of older populations, causing younger populations of ages $<$ 2 Gyr to have steeper observed gradients. They found that radial migration is more effective in the radial range 5 to 11 kpc and small beyond 12 kpc. As a result, steeper radial metallicity gradients produced by the younger stars in the radial range 5 to 11 kpc persist for a few Gyr and eventually washed out at latter times when the guiding radius of stars change and migrate to other locations in the disc.

From the analysis of high-quality spectra of FGK stars in the solar vicinity, Haywood et al. (2013) argue for a lack radial migration in the sense of churning and suggested that the orbital eccentricity of stars caused by blurring alone would explain the observable properties in the solar vicinity.
Along these lines we suggest that the relatively large eccentricity of thick disc OCs (Table 16) is sufficient to explain their present locations in the outer disc, and the striking chemical similarity between these OCs further strengthens that they were formed at around 11 kpc in the Galactic disc from a medium that was chemically homogeneous for atleast a few kiloparsecs.

\section{Summary and conclusions}
One aim of our studies of the chemical compositions of red giants in OCs is to provide insights into Galactic chemical evolution (GCE).
Based on the high resolution spectra of red giants in twelve OCs and standard LTE differential abundance analysis relative to the Sun, we presented stellar parameters and the first estimates of chemical abundances of 26 species from Na to Eu sampling all the major processes of stellar nucleosynthesis -- $\alpha$ and $r$-process elements synthesized in Type II supernovae from massive stars, iron group elements synthesized in Type Ia supernovae, and the $s$-process elements synthesized in asymptotic-giant-branch (AGB) stars. We combined with our total sample of 28 clusters from this and previous papers with a sample of 51 OCs drawn from the literature whose chemical abundances and Galactocentric distances were remeasured to establish a common abundance and distance scales, respectively. This compilation provides a homogenised sample of 79 OCs spanning a narrow range in metallicity of $\sim$ $-$0.5 to 0.3 dex but a range of a few Myr to 9 Gyr in ages and 5.0 to 24.0 kpc in R$_{\rm gc}$. Our sample alone constitute about 35\% of OCs explored so far with high-resolution spectroscopy and we are the first to provide a homogeneous high-resolution abundance analysis of elements from Na to Eu for twenty-eight OCs in the radial range R$_{\rm gc}$=\,7.7 to 11.3 kpc; only a few OCs in the literature present abundance estimates for the heavy elements from Y to Eu.

Following the kinematic criteria, we assigned OCs to the thin disc, thick disc or halo stellar populations and examined the age-abundance relations and the variation of [Fe/H] as a function of R$_{\rm gc}$ in the Galactic disc. We showed that the resolution of the lack of age$-$metallicity relation for OCs, as noted previously by other studies, lies in the incomplete coverage in metallicity compared to that of field stars which cover the range of $-$1.0 to $+$0.4 dex in metallicity and an age of a few Myr to 13 Gyr. 

Using our homogeneous sample of 79 OCs, we confirm the results of previous studies that OCs present a constant steep decline of metallicity out to R$_{\rm gc}$ of 12 kpc and a further flattening out to the entire radial extent of the Galactic disc. But our analysis is the first to demonstrate clearly that such bimodality accompanied by a sudden change in the slope of radial metallicity distribution of OCs at 12 kpc arise from the selection 
effects; at R$_{\rm\,gc}<\,$12 kpc, all the sampled clusters lie close to the Galactic midplane ($\lvert\,z\rvert<\,$0.5 kpc), younger than 1.5 Gyr with kinematics typical of thin disc and are constituting a constant steep decline of [Fe/H] with R$_{\rm\,gc}$ while the OCs populating the disc beyond 12 kpc are older with ages from 1.0 to 8.0 Gyr, metal-poor by [Fe/H] $<-$0.2 dex with thick disc kinematics and located away from the midplane (0.5$\,<\lvert\,z\rvert<\,$2.5 kpc) and constitute a shallow gradient over the entire radial extent of the Galactic disc followed by a change of slope at 12 kpc.

We further compared the gradients traced by OCs with that of field stars, Cepheids and with chemodynamical model predictions. We demonstrated clearly that the OCs, field stars and Cepheids yield consistent radial gradients if the comparison samples are drawn from the similar vertical slices. The radial metallicity gradient traced by the field stars and Cepheids lying close to the midplane ($\lvert\,z\rvert<\,$0.5 kpc) are steeper while that measured for field stars located away from the midplane (0.5$\,<\lvert\,z\rvert<\,$2.0 kpc) are shallower over the entire radial extent of the disc and are comparable to similar gradients measured for the sample of OCs. The chemodynamical models of Minchev et al. (2013) produce a steep gradient 
for samples of stars younger than 2 Gyr and a flat trend for stars older than 2 Gyr throughout the radial range 5 to 16 kpc of the disc. In their simulations, a signature of flattening of the radial gradient arise naturally due to the radial mixing of stars but the young population is hardly affected as the radial mixing operates at a slow pace. As a result, the younger populations are expected to show a steeper gradient than that of the older ones. Taking into account the affects of radial mixing, these simulations fairly predict a steep gradient of $-$0.058 dex kpc$^{-1}$ (R$_{\rm\,gc}=$6$-$11 kpc) for the stellar populations younger than 2 Gyr and the shallow gradient of old populations which are in fair agreement with such gradients observed for the tracers such as OCs, field stars and Cepheids.

Finally, we demonstrated through the computation of birthplaces of OCs that the sample of clusters (ages$<$ 1.5 Gyr and R$_{\rm\,gc}<\,$12 kpc) constituting a steep radial metallicity gradient of slope $-$0.052$\pm$0.011 dex kpc$^{-1}$ have circular orbits with birthplaces very close their present locations in the Galactic midplane. In contrast, the older clusters populating the outer disc (R$_{\rm\,gc}>\,$12 kpc) with a measured shallow slope of $-$0.015$\pm$0.007 dex kpc$^{-1}$ have relatively eccentric orbits but with birthplaces around R$_{\rm\,gc}$ of 11 kpc close to the midplane. Our analysis is the first to demonstrate clearly that the orbital eccentricity of all but the three outer disc clusters Be 29, Be 75 and Saurer 1 has taken them to present locations in the Galactic disc from their birthplaces inward of 12 kpc from a medium enriched with very similar metallicity. Moreover, these outer disc clusters also make large excursions away from the Galactic midplane, a pleasing result in support of their survival for longer periods of time. The older, distant clusters Be 29 and Saurer 1 at a height of 2 kpc and 1.7 kpc, respectively, are of extra-galactic origin while Be 75 may be an old genuine Galactic cluster on a perturbed orbit or the astrometric data employed in orbit calculations might be erroneous and require further scrutiny.

In the spirit of speculation to encourage further spectroscopic analyses, the suggestion is made that the variation of metallicity of OCs with Galactocentric distance is a linear function with a constant negative slope over the entire radial extent of the disc but close to the Galactic midplane. The radial gradients become flatter as one moves away from the Galactic midplane and for the older OCs. To test such speculations, it would be useful to explore the chemical content of as many clusters as possible including the young and old OCs covering a wide range in R$_{\rm gc}$ from 5 to 18 kpc in the disc and a scale height of 0.0 to 3 kpc above the midplane and then to compare the abundance gradients of OCs with those derived from the field stars and Cepheids.

\vskip1ex 
{\bf Acknowledgements:}

We thank the anonymous referee for comments which have improved the Paper. We are grateful to the McDonald Observatory's Time Allocation Committee for granting us observing time for this project. DLL wishes to thank the Robert A. Welch Foundation of Houston, Texas for support through grant F-634. ABSR thanks Dr. Zhen-Yu Wu for generously providing his code for the integration of open cluster orbits.

This research has made use of the WEBDA database, operated at the Institute for Astronomy of the University of Vienna and the NASA ADS, USA. This research has also made use of Aladin. This publication makes use of data products from the Two Micron All Sky Survey, which is a joint project of the University of Massachusetts and the Infrared Processing and Analysis Center/California Institute of Technology, funded by the National Aeronautics and Space Administration (NASA) and the National Science Foundation (NSF).

\end{document}